# Giant Negative Linear Compressibility in Orthorhombic Copper Cyanide

Swayam Kesari[1], Alka B. Garg[2,3], Nilesh P. Salke[1,4,*] and Rekha Rao[1,3,*]

[1]Solid State Physics Division, [2]High Pressure and Synchrotron Radiation Physics Division, Bhabha Atomic Research Centre, Mumbai 400085, India.

[3] Homi Bhabha National Institute, Anushaktinagar, Mumbai 400094, India

[4] Department of Physics, University of Illinois Chicago, Chicago 60607, USA

## Abstract

Most reported negative linear compressibility (NLC) materials exhibit either a small NLC over a large pressure range or a high NLC over a very small pressure range. Here, we report the remarkable discovery of giant NLC in the low-temperature form of CuCN (LT-CuCN) over an unusually large pressure range. High-pressure XRD studies on LT-CuCN observed the NLC of -20.5 TPa$^{-1}$ along the *a*-axis at zero pressure, and the ambient orthorhombic phase remained stable up to 9.8 GPa. Pressure and temperature-dependent Raman studies identified the phonon vibrations responsible for NLC and negative thermal expansion (NTE).

[*]*Corresponding author*
rekhar@barc.gov.in
nilesh@uic.edu

**Main Text**

Under hydrostatic compression, most materials usually contract in all crystal directions, showing positive linear compressibility (PLC). However, in rare cases, some materials expand along one or more crystal directions, demonstrating negative linear compressibility (NLC) or negative area compressibility (NAC). NLC is a relatively very weak phenomenon compared to the typical PLC in materials.[1,2] Very few materials have been reported to have strong NLC; most are found in framework structures. The maximum NLC observed till now in a material is -76 TPa$^{-1}$($K_c$) in Ag$_3$[Co(CN)$_6$] along the *c*-direction and only up to 0.19 GPa of pressure[3], above which it undergoes a structural phase transition[3-5]. Another notable example is the giant NLC of -42(5) TPa$^{-1}$ in Zinc dicyanoaurate Zn[Au(CN)$_2$] along the *c*-direction but only up to 1.8 GPa of pressure, above which it shows structural transition and persisted NLC of -6(3) TPa$^{-1}$ up to 14.2 GPa of pressure.[6-8] Among other NLC materials with a large range, most have very small NLC values; one notable example is β-MnO$_2$, which shows an NLC of merely -0.16 TPa$^{-1}$ for a pressure range from 0.3 to 29.3 GPa.[9] Most of the materials discovered either have very weak NLC or have a very short range of pressures. Strong NLC materials show structural instability; hence, observing large NLCs for a large pressure range has been challenging.

Anisotropic thermal expansion exhibited by cyanides has been of interest following reports of NLC and colossal negative thermal expansion (NTE) in Ag$_3$Co(CN)$_6$.[3-5,10] Several cyanides such as Ag$_3$[Co(CN)$_6$], KMn[Ag(CN)$_2$]$_3$, KCd[Ag(CN)$_2$]$_3$ and Zn[Au(CN)$_2$]$_2$ are reported to show both NLC & NTE behaviours simultaneously.[3,4,6-8,11-13] Interestingly, most of these materials show structural phase transition under high pressure by a maximum of about 3 GPa, above which either they amorphized, decomposed, or transitioned into new structures that cease to show NLC and/or NTE. Here, it is quite evident that observing strong NLC for a large pressure range is not possible. Besides, cyanides are also interesting due to their unusual properties, like pressure-induced amorphization (PIA) and pressure-induced polymerization (PIP). It is important to investigate the fundamental physics behind anomalous NLC and NTE properties in designing and developing composite materials. NLC is also an attractive mechanical property, with a key application being the development of effectively incompressible optical materials.[1,2] Metal cyanides of the type *M*CN (*M*=Au, Ag, and Cu) form linear chain structures and show one-dimensional negative thermal expansion along the chain direction with overall positive thermal expansion.[14] However, there are no reports of NLC in these kinds of materials. CuCN crystallizes into orthorhombic or hexagonal phases at ambient conditions depending on the synthesis condition, named low-temperature form as LT-CuCN and high-temperature form as HT-CuCN phases, respectively, which are disordered.[15,16] LT-CuCN crystallizes in an orthorhombic structure with space group $C222_1$ with Z=4, and HT-CuCN crystallizes in a hexagonal phase belonging to *R3m* with Z=3. The LT-CuCN phase converts irreversibly to the HT-CuCN phase at 563 K.[15,16] Of all the *M*CN (*M*=Cu, Ag, and Au), LT-CuCN has the largest coefficient of NTE along the chain direction. For LT-CuCN, the thermal expansion coefficient along the *a*-direction is -53.8 × $10^{-6}$ $K^{-1}$, and for HT-CuCN, it is -27.9 × $10^{-6}$ $K^{-1}$ in the *c*-direction.[14]

Like other cyanides where NTE co-existed with strong NLC, it is expected to have strong NLC in LT-CuCN. Hence, the structural properties of LT-CuCN were investigated using synchrotron x-ray diffraction at high pressures. Interestingly, strong NLC is found in LT-CuCN, which remained in the ambient phase over large pressure range. Further, the vibrational properties of LT-CuCN have been investigated by pressure- and temperature-dependent Raman measurement. Soft modes contributing to NTE in LT-CuCN are identified. In addition, Raman spectroscopy has provided valuable information about the anharmonicity of the modes, structural stability and the existence of new phases.

LT-CuCN crystallizes in space group $C222_1$ with Z=4, which contains wave-like 1-D chains in which two-coordinated Cu are bridged by site-disordered CN.[15,16] Fig. 1(a) depicts the crystal structure of LT-CuCN. The structure consists of infinite CuCN chains containing five crystallographically distinct Cu atoms with nine CuCN units. Fig. 1(b) represents the x-ray diffraction pattern of LT-CuCN at selected pressures up to 20.2 GPa (See supplementary Material for experimental details). The XRD pattern at closer pressure intervals are also given in the Fig. S1 (see Supplementary Material). The lattice parameters at several pressures up to 9.8 GPa were refined. The Rietveld [17] refined XRD pattern at **(a)** 0.4, **(b)** 0.5, **(c)** 0.8, **(d)** 1.4, **(e)** 2.1, **(f)** 3.4 **(g)** 4.8, **(h)** 6.4, **(i)** 7.9 and **(j)** 9.8 GPa pressures are shown in supplementary Fig. S2 (see Supplementary Material), and the variation of lattice parameters and unit cell volume of LT-CuCN with pressure is shown in Fig. 1(c)-(e), respectively. The refined atom positions (Cu positions were refined while the C and N positions were kept fixed at the reported position [15]) at all the pressure data are given in the Table S1-10(see Supplementary Material).

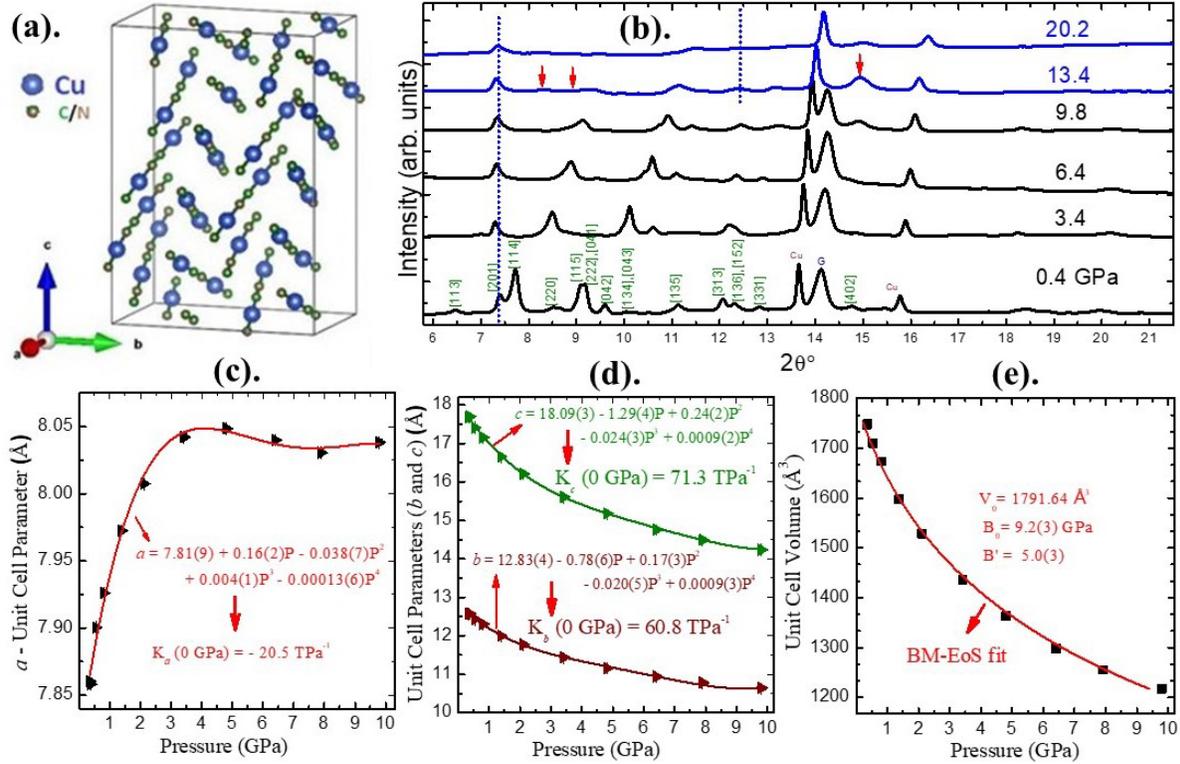

**FIGURE 1:** X-ray diffraction studies of LT-CuCN at high pressures **(a).** Crystal Structure of LT-CuCN phase with space group $C222_1$. **(b).** X-ray diffraction pattern of LT-CuCN at various high pressures ($\lambda$ = 0.4957 Å). Peaks labeled with Cu indicate reflection peaks for copper, which is used as a pressure marker. Peaks labeled with G indicate the reflection peaks of the gasket material. Vertical dotted blue lines are a guide to see the anomalous movement of the reflection peak towards the lower 2θ angles with increasing pressures. The black curve indicates the ambient orthorhombic phase and the blue curve indicates the new high-pressure phase. **(c).** Pressure dependence of *a*- unit cell parameter. Black symbols show experimental *a*-unit cell parameters. The red curve represents polynomial fit. **(d).** Variation of *b*, *c*-unit cell parameters with pressure. The green and brown curves represent polynomial fits to *b* and *c*-unit parameter data. **(e).** Variation of unit cell volume with pressure. The red curve represents BM-EoS fitting. Error bars in **(c)-(e)** are within the symbol size.

The high-intensity reflection peak appears at 7.7°, marked as [114], moves towards a higher angle under pressure, indicative of lattice compression. Interestingly, the anomalous shift under high pressure is observed on the reflection peak at 7.3°, marked as [201] crystallographic plane. The variation of inter-planar spacing of prominent lattice planes of LT-CuCN in its orthorhombic structure is shown in supplementary Fig. S3 (see Supplementary Material) and are also given in the Table S11 (see Supplementary Material). The inter-planar separation for [201] crystallographic plane indicates anomalous behavior compared to other planes. This unusual behavior is an indication of NLC. This expansion may be a precursor to a structural phase transition at higher pressures. In addition to this anomalous shift of the diffraction peak, the XRD pattern remains qualitatively similar up to 9.8 GPa. The XRD data analyzed using Rietveld refinement with ambient orthorhombic $C222_1$ structure up to 9.8 GPa. Around 11.6 GPa, a few new reflection peaks appeared at 8.4º and 14.9 º, also marked in Fig. 1(b), which couldn't be fitted with ambient structure. Also, the major reflection peak observed at 7.7º around 0.4 GPa disappears completely by 18 GPa. This observation indicates structural phase transition in LT-CuCN above 9.8 GPa of pressure. The variation of unit cell volume with pressure shown in Fig. 1(e) is fitted with the third-order Birch Murnaghan Equation of State (BM-EoS), and the zero pressure bulk modulus of LT-CuCN is estimated to be $B_0 = 9.2(3)$ GPa. The obtained value of bulk modulus suggests that the compound LT-CuCN is very soft and in the range of many of the observed NLC cyanides. The variation of lattice parameters of LT-CuCN with pressure are shown in Fig. 1(c)-1(d). It is very important to note here that the lattice parameters "$b$" and "$c$" showed usual contraction under compression i.e. positive compressibility. In contrast, lattice parameter "$a$" showed interesting, unusual expansion under pressure, i.e. NLC in its orthorhombic phase. The orthorhombic LT-CuCN remains structurally stable and is NLC up to 9.8 GPa. The zero pressure axial compressibility along '$a$', '$b$', and '$c$' are estimated to be -20.5, 60.8, and 71.3 TPa$^{-1}$ ($K_a$, $K_b$, $K_c$), respectively. This linear compressibility along the three directions are estimated using polynomial fits, as shown in Fig. 1(c) and Fig. 1(d). The bulk modulus calculated using $\frac{1}{K_a+K_b+K_c}$ is 8.9 GPa, close to the value obtained by fitting experimental volume data to B-M EoS, see Fig. 1(e). Cairns and Goodwin reviewed the use of linear fitting, polynomial fitting, and empirical model [1,6] to estimate the axial compressibility of NLC materials from the pressure variable lattice parameter experiments. The same empirical model is also used in the program PASCAL [18], from which the compressibilities for several systems in the literature are obtained. Our data did not fit well with the linear and with the empirical model (See Fig. S4(a) & S4(b) in Supplementary material) given in [1,6], hence we used polynomial fit in our studies (See Fig. S4(c) in the Supplementary Material for comparison to Linear or empirical model fit). We compared the obtained compressibility value of LT-CuCN with those already discovered NLC cyanides Ag$_3$[Co(CN)$_6$], KMn[Ag(CN)$_2$]$_3$, KCd[Ag(CN)$_2$]$_3$, Zn[Au(CN)$_2$]$_2$ and Eu[Ag(CN)$_2$]$_3$·3H$_2$O. In these cyanides, compressibility values were reported using linear fit or from empirical model. To make a fair comparison, we fitted their reported pressure-dependent lattice parameters using polynomials and estimated the NLC at zero pressure. The polynomial fitting to their reported data are shown in the Fig. S5(a) to 5(f). The zero pressure compressibility along their NLC directions are compared in the Table S12. The pressure variation of linear compressibility for these cyanides are calculated and shown in the Fig. 2. Interestingly, the magnitude of K$_a$ for LT-CuCN (K$_a$ (0 GPa) = -20.5 TPa$^{-1}$) initially decreases and reaches a minimum value of -1.8 TPa$^{-1}$ at 4.81 GPa, then shows an increment in K$_a$ under higher pressures to again to a larger value of -9.4 TPa$^{-1}$ at 9.8 GPa. The mean NLC value for LT-CuCN over 0-9.8 GPa pressure range is estimated to -10.2 TPa$^{-1}$. These studies prove that LT-CuCN has been discovered to be the first cyanide to have a giant NLC compressibility value for a remarkably large pressure range reported so far in the ambient structure. Interestingly, after the transition, the [201] reflection peak, shifting towards a lower 2θ angle in the orthorhombic phase, indicative of anomalous expansion under compression, began to shift towards a higher 2θ angle with compression. Simultaneously, another reflection peak at 12.4° around 11.6 GPa shifted towards lower angles at high pressure, indicating that the NLC behavior still exists in the new phase. (see Supplementary Fig. S1 in Supplementary Material).

Observing giant NLC in LT-CuCN over a large pressure range (at least up to 9.8 GPa) is rare and unique. In the development of effectively incompressible optical materials, NLC will have applications.[1,2] The literature speculates that stronger NLC can be found in framework structures showing very anisotropic NTE, although both can be independent phenomena.[1] It is also interesting to note that the discovered NLC and reported NTE in LT-CuCN are observed along the same *a*-lattice parameter. To our knowledge, only a limited number of NLC cyanides such as Ag$_3$[Co(CN)$_6$], KMn[Ag(CN)$_2$]$_3$, KCd[Ag(CN)$_2$]$_3$ and Zn[Au(CN)$_2$]$_2$,[3,4,6-8,11-13] were reported to show both NLC & NTE behaviours in their ambient phases. But these materials show structural phase transition at high pressure of 0.19 [3], 1.8 [12,13], 0.5 [11] and 2.8 [6] GPa, respectively. Recently Eu[Ag(CN)$_2$]$_3$·3H$_2$O is reported to show NLC mean compressibility of -4.2(1) TPa$^{-1}$ up to 8.2 GPa [19] which was obtained using strain tensor. The mean compressibility for Eu[Ag(CN)$_2$]3·3H$_2$O using polynomial will be -6.2 TPa$^{-1}$ for the pressure range 0-8.2 GPa.

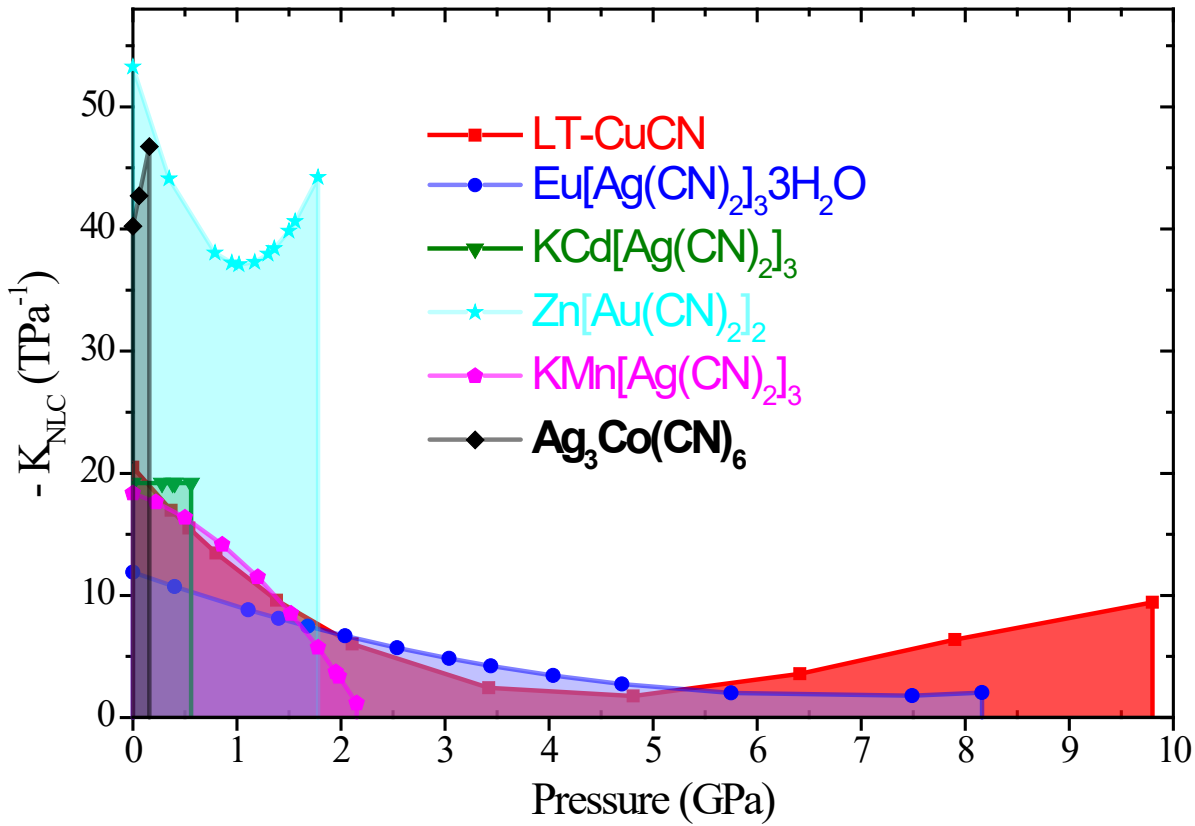

**FIGURE 2**: Pressure variation of Linear compressibility (along the NLC direction) within pressure range of structural stability of their ambient phases of LT-CuCN, Ag$_3$[Co(CN)$_6$], KCd[Ag(CN)$_2$]$_3$, Zn[Au(CN)$_2$]$_2$, KMn[Ag(CN)$_2$]$_3$ and Eu[Ag(CN)$_2$]3·3H$_2$O estimated after fitting polynomials to their lattice parameter data. For polynomial fitting see Fig. S5(a)-(f) (see Supplementary Material).

In KMn[Ag(CN)$_2$]$_3$, the NLC was reported to have a mean value of $K_c$ = -12.0(8) TPa$^{-1}$ using empirical model fit, which will become $K_c$ = -11.2 TPa$^{-1}$ (mean value over the experimental data) using polynomial fit. The ambient structure of KMn[Ag(CN)$_2$]$_3$ changes at 2.8 GPa, followed by PIA at 13.5 GPa.[12,13] In KCd[Ag(CN)$_2$]$_3$, the NLC was reported to have a mean value of $K_c$ = -21(2) TPa$^{-1}$ using linear fit in its first phase followed by three phase transitions (at 0.5, 2 & 3 GPa) in which all three phases show NLC behavior.[11] In Zn[Au(CN)$_2$]$_2$, the NLC was observed with a mean value of $K_c$= -42(5) TPa$^{-1}$ using the empirical model fit, which will be - 40.7 TPa$^{-1}$ (mean value over the experimental data) using polynomial fit. The phase transition in Zn[Au(CN)$_2$]$_2$ appears around 1.8 GPa.[6-8] For Ag$_3$[Co(CN)$_6$], the mean value of K$_c$ = -76 TPa$^{-1}$ was reported using a linear fit which is also same as at zero pressure.[3] DFT [20] and DFT+D2 [21] studies on Ag$_3$[Co(CN)$_6$] indicated zero pressure NLC value at 0 K of about K$_c$ = -18.8 and K$_c$ = -21 TPa$^{-1}$, respectively. A large underestimation of NLC value in theoretical calculation was assigned to the considerable softening of the material on heating [21]. We found that the reported pressure-dependent $c$-lattice parameter data for Ag3[Co(CN)6] (within 0.19 GPa) [3] fits well with the second-order polynomial compared to linear fit, as shown in Fig. S5(c) ) (see Supplementary Material) with zero pressire K$_c$ = - 40.2 TPa$^{-1}$. NLC obtained with polynomial fit is reasonably consistent with theoretically calculated NLC. [20,21] This indicates that proper care should be taken to estimate NLC by ensuring the best fit. The axial compressibility and bulk moduli reported for different cyanides are presented in Supplementary Table S12 (see Supplementary Material) for comparison. We would like to highlight that the mean value of NLC in LT-CuCN is of the same order of magnitude as NLC at zero pressure. However, we believe that the average or mean value does not fully capture the true nature of NLC. Closer data points near the pressure where the NLC is strongest can skew the mean toward higher values. Therefore, specifying the compressibility at zero pressure or at a given pressure is more appropriate than relying on average values. Interestingly after polynomial fitting the NLC value at zero pressure varies as K$_c$(Zn[Au(CN)$_2$]$_2$) > K$_c$(Ag$_3$[Co(CN)$_6$]) > K$_a$(LT-CuCN)> K$_c$(KCd[Ag(CN)$_2$]$_3$) > K$_c$ (KMn[Ag(CN)$_2$]$_3$) > K$_c$(Eu[Ag(CN)$_2$]$_3$·3H$_2$O) which is shown in the Fig. S5 and the Table S12 (see Supplementary Materials). Observation of NLC and structural stability of LT-CuCN in the orthorhombic phase over a large pressure range (up to about 9.8 GPa) makes it a potential candidate for technological applications.

Further, pressure-dependent Raman investigations are also carried out to understand the behavior of phonons for this discovered NLC material. Raman spectra of LT-CuCN in three frequency regions at ambient conditions are shown in Supplementary Fig. S6 (see Supplementary Material). Raman spectra could be deconvoluted into fifteen Lorentzian bands. Some of the Raman mode frequencies in the bending and stretching region match well with the previously reported Raman spectra of LT-CuCN [15,22]. The Raman bands below 50 cm$^{-1}$ are being reported for the first time here, clearly showing two lattice modes at 20 and 30 cm$^{-1}$. The modes assignments are taken from the literature.[22] The earlier high-pressure Raman study on LT-CuCN was reported only up to 3.7 GPa,[22] which gave limited information. Our interest is also to probe the low-frequency modes, representing lattice vibrations, which are generally held responsible for NTE in other cyanides [23,24], and the high-pressure vibrational behavior beyond the NLC range discovered in LT-CuCN. Fig. 3(a)-(c) shows the Raman spectra of LT-CuCN at various pressures in the three-frequency range (a). 15-75 cm$^{-1}$, (b). 75-675 cm$^{-1}$ and (c). 2050-2250 cm$^{-1}$. Both lattice modes at 20 and 30 cm$^{-1}$ display stiffening with pressure; therefore, they do not contribute to the NTE behavior in LT-CuCN (Fig 3a and 3d).

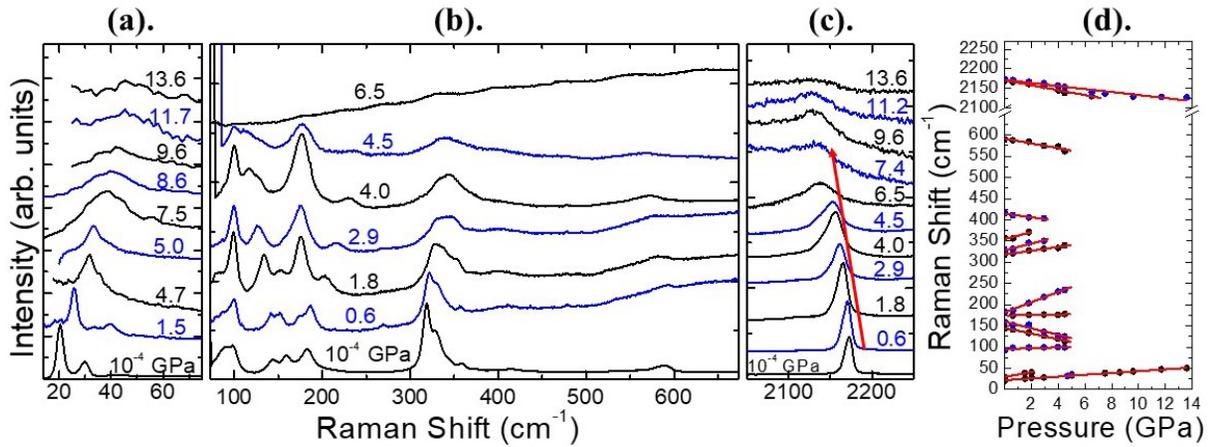

**FIGURE 3:** Raman spectroscopic studies of LT-CuCN at high pressures. **(a)**, **(b)** and **(c)** represent Raman spectra of LT-CuCN at frequency ranges 15-75 cm$^{-1}$, 75-675 cm$^{-1}$, and 2050-2250 cm$^{-1}$ respectively. **(d).** Variation of Raman mode frequencies with pressure. Solid lines indicate a linear fit to the data.

New Raman mode appears around 175 cm$^{-1}$ at 0.6 GPa, showing a continuous increase in intensity with pressure; see Fig. 3(b) and 3(d). It is interesting to note that the modes at 158, 580, and 2170 cm$^{-1}$ assigned to Cu-C-N-Cu bending, Cu-C/N stretching, and the C≡N stretching modes, respectively, show unusual softening with pressure as can be seen in Fig. 3(b), 3(c), and 3(d). This softening of the modes reveals abnormal elongation of C≡N bonds. An increase in inter-planar separation with pressure for [201] plane that is associated with NLC observed in high-pressure XRD can be explained by the softening of the vibrational modes in the Raman spectrum since the CuCN chain is across this plane, and overall chain length could be increasing with pressure. Mechanisms for observing the NLC property have been classified into four categories viz, (i) NLC as a consequence of ferroelastic phase transition, (ii) NLC driven by correlated polyhedral tilts, (iii) NLC in helical systems, and (iv) NLC due to framework hinging in framework materials.[1] The present NLC material LT-CuCN has an orthorhombic structure containing periodic chains of Cu-C-N-Cu, mostly aligned along *a*-direction. Under hydrostatic pressure, we observed abnormal elongation of the interplanar spacing of the [201] plane from the XRD investigations and softening of the CN stretching frequency from the Raman investigation. This observation indirectly indicates that the chain length will increase under hydrostatic pressure. From XRD, it is difficult to refine the C and N positions to give a detailed picture of the overall increase in the chain length. In the refinement of the XRD data, the Cu positions were moved, keeping C and N positions fixed to the reported positions [15].

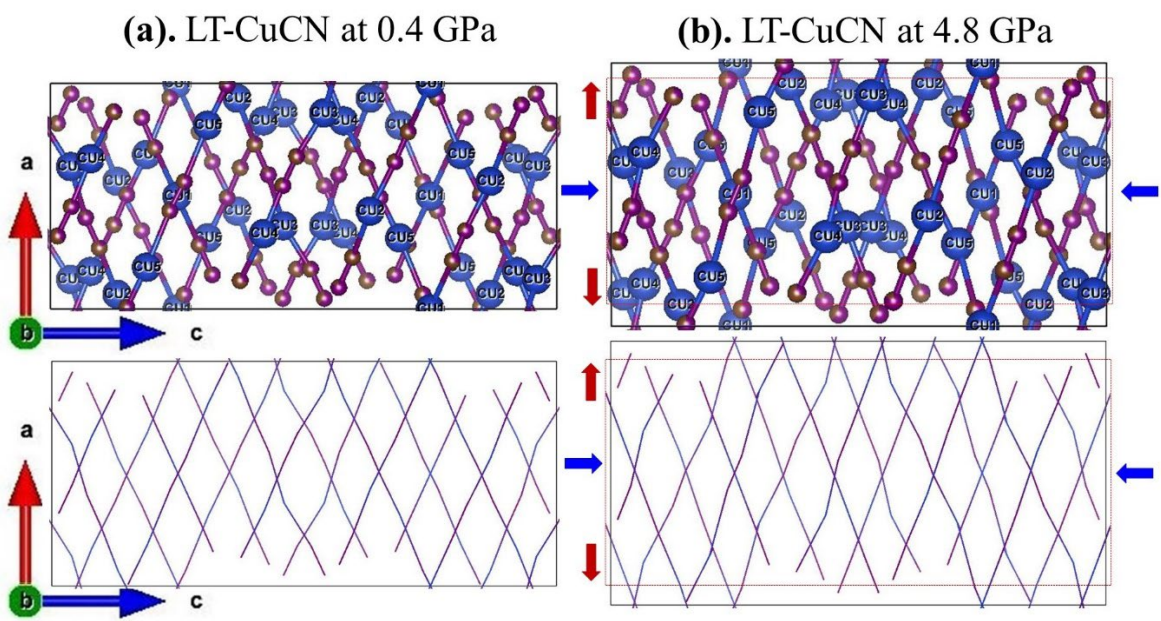

**FIGURE 4:** Structure of LT-CuCN at **(a).** 0.4 and **(b).** 4.8 GPa viewed in the ac plane. Elongation along a-axis under high pressure is observed due to hinging of the wine-rack structure.

Fig. 4(a) and 4(b) indicate the arrangement of Cu-CN-Cu chains in the orthorhombic structure when viewed in the *ac* plane at 0.4 and 4.8 GPa. In Fig. S7(a) to (j) (See Supplementary Material), the structure of LT-CuCN viewed in the *ac* plane is given for all the pressure data for better visualization. The structure of LT-CuCN in the *ac* plane is a wine rack structure. Hinging of the wine rack network under high pressure is responsible for the unusual elongation of the *a*-lattice parameter. This mechanism explains the NLC observed in the present orthorhombic LT-CuCN. A similar mechanism also has been observed for Ag$_3$[Co(CN)$_6$] [3], KMn[Ag(CN)$_2$]$_3$ [13], Zn[Au(CN)$_2$]$_2$ [6] and KCd[Ag(CN)$_2$]$_3$ [11] etc.

Fig. 3(d) shows the variation of Raman mode frequencies with pressure, indicating that some of the modes show softening behavior in addition to the hardening of most of the other modes. The softening of the C-N stretching mode is marked with an arrow in Fig. 3(c). Phonon modes with negative isothermal Grüneisen parameters mode are called as soft modes. Observation of soft modes indicates that the phonon frequency is decreasing and approaching zero frequency, indicating thermodynamical instability, which may be responsible for a structural phase transition in a crystalline material [25]. The thermal expansion coefficient given as $\alpha_V = \frac{1}{B_0 V}\sum_i C_i \gamma_{iT}$ where $C_i$ is the isothermal specific heat, $V$ the unit cell volume, $B_0$ the bulk modulus, and $\gamma_{iT}$ is the isothermal mode Grüneisen parameter for the i$^{th}$ mode. This expression shows that the soft modes contribute negatively to the thermal expansion coefficient as they have negative $\gamma_{iT}$. The variation of Raman mode frequencies of LT-CuCN with pressure are fitted linearly, and the slopes are given in Supplementary Table S13 in Supplementary Material. From Fig. 3(a)-(c), it is also clear that the intensity of all the Raman modes is reduced. The disappearance of the Raman spectra in the middle-frequency region above 7 GPa is observed. In addition, ultra-low frequency and the C-N stretching frequency modes remained visible (although weak in intensity) with almost similar pressure-frequency dependence up to 13.6 GPa see Fig. 3(a) and 3(c). Above 13.6 GPa, the whole of the Raman spectra disappeared. This behavior confirms that the crystal structure of LT-CuCN is changing, as observed in high-pressure XRD discussed in the previous section. Since Raman spectroscopy is a local probe that has detected the transition starting around 7 GPa and completing above 13.6 GPa. However, there was no significant broadening of the Raman modes before the spectra disappears. Hence, amorphization is ruled out, which is also consistent with the XRD results. The complete disappearance of the Raman spectra indicates the appearance of a new crystal/electronic structure at high pressure, and the continuous softening of the C≡N stretching mode could also be a precursor to a probable dissociation. However, when released from 20.2 GPa, the signatures of the ambient LT-CuCN phase were observed in Raman and XRD measurement, indicating the reversible nature of the phase transition (see Supplementary Material Fig. S8 and Fig. S9).

Thermal expansion is an experimental observation of anharmonicity in the interatomic potential energy. Anharmonicity indicates the contribution of the cubic and quartic terms in the potential energy expression, which results from three phonons and four phonons decay processes, respectively.[26,27] The anharmonicity analysis is carried out after combining high-pressure- and temperature-dependent investigations of vibrational modes of LT-CuCN. For an isotropic system, the variation of a phonon mode frequency with temperature is in general, given as [26]

$$\frac{1}{\omega_i}\left(\frac{\partial \omega_i}{\partial T}\right)_P = -\alpha_V \gamma_{iT} + \frac{1}{\omega_i}\left(\frac{\partial \omega_i}{\partial T}\right)_V \quad\ldots\ldots\ldots\ldots\ldots\ldots(1)$$

where $\omega_i$ is the wavenumber of $i^{th}$ phonon mode, $\alpha_V$ is the coefficient of volumetric thermal expansion and $\gamma_{iT}$ is the isothermal mode Grüneisen parameter. The term on the left-hand side in equation (1) is total anharmonicity. The first term on the right-hand side in equation (1) is the implicit anharmonicity and the second term on the right-hand side represents the true anharmonicity and has purely temperature effect due to phonon-phonon decay, which is also known as explicit anharmonicity.[26-28] The Raman spectra of LT-CuCN at various temperatures are shown in Fig. S10(a)-(c) (see Supplementary Material). The variation of all the Raman mode frequencies with temperature is shown in Supplementary Fig. S10(d) (see Supplementary Material). With temperature, all the observed Raman modes show usual behavior, which is a decrease in frequency with increasing temperature, and no anomalous modes have been observed. The variation of Raman mode frequencies of LT-CuCN with temperature is fitted linear, and the slopes are given in Supplementary Table S14 (see Supplementary Material). The three anharmonicity components are estimated using the expression (1) and are given in Supplementary Table S14 (see Supplementary Material). All the observed modes show significant contributions due to the three phonon decay processes in anharmonicities. It is also observed that the total anharmonicity of the low-frequency modes is large, especially the external modes.

In conclusion, high-pressure synchrotron X-ray diffraction and Raman spectroscopic investigations were carried out on LT-CuCN up to 20.2 and 13.6 GPa, respectively. The compound LT-CuCN, known to show anisotropic thermal expansion, is discovered to have large anisotropic compressibility, showing NLC along *a*-axis over a wide pressure range due to the hinging of wine-rack structure of LT-CuCN. The negative linear compressibility obtained in LT-CuCN for the pressure range 0-9.8 GPa (mean value of -10.2 TPa$^{-1}$) is found to be larger than with the recently discovered NLC compound Eu[Ag(CN)$_2$]$_3$·3H$_2$O for the range 0-8.2 GPa[19]. Some of the Raman modes in LT-CuCN show softening under high pressure. Changes observed in Raman spectra above 13.6 GPa are consistent with the appearance/disappearance of reflection peaks in the XRD pattern above 9.8 GPa, which indicates a structural phase transition.

## ACKNOWLEDGMENTS

S.K., A.B.G., and R.R. acknowledge DST, India, for the financial support for the work done at the Elettra synchrotron source with proposal number 20165031. The authors thank Dr. S. N. Achary of the Chemistry Division, Bhabha Atomic Research Centre, for providing XRD data of LT-CuCN at ambient conditions. N.P.S. acknowledges support by DOE-SC (DE-SC0020340), DOE-NNSA (DE-NA0004153, CDAC), and NSF (DMR-2119308). N.P.S. acknowledges the support and encouragement of Prof. Russell J. Hemley.

## REFERENCES

[1]  A. B. Cairns and A. L. Goodwin, Physical Chemistry Chemical Physics **17**, 20449 (2015).
[2]  R. H. Baughman, S. Stafström, C. Cui, and S. O. Dantas, Science **279**, 1522 (1998).
[3]  A. L. Goodwin, D. A. Keen, and M. G. Tucker, Proceedings of the National Academy of Sciences **105**, 18708 (2008).
[4]  L. Wang, C. Wang, H. Luo, and Y. Sun, The Journal of Physical Chemistry C **121**, 333 (2017).
[5]  R. Rao, S. N. Achary, A. K. Tyagi, and T. Sakuntala, Physical Review B **84**, 054107 (2011).
[6]  A. B. Cairns *et al.*, Nature Materials **12**, 212 (2013).
[7]  M. K. Gupta, B. Singh, R. Mittal, M. Zbiri, A. B. Cairns, A. L. Goodwin, H. Schober, and S. L. Chaplot, Physical Review B **96**, 214303 (2017).
[8]  L. Wang, H. Luo, S. Deng, Y. Sun, and C. Wang, Inorganic Chemistry **56**, 15101 (2017).
[9]  J. Haines, J. M. Léger, and S. Hoyau, Journal of Physics and Chemistry of Solids **56**, 965 (1995).
[10] A. L. Goodwin, M. Calleja, M. J. Conterio, M. T. Dove, J. S. O. Evans, D. A. Keen, L. Peters, and M. G. Tucker, Science **319**, 794 (2008).

[11]     A. B. Cairns et al., The Journal of Physical Chemistry C **124**, 6896 (2020).
[12]     K. Kamali, C. Ravi, T. R. Ravindran, R. M. Sarguna, T. N. Sairam, and G. Kaur, The Journal of Physical Chemistry C **117**, 25704 (2013).
[13]     A. B. Cairns, A. L. Thompson, M. G. Tucker, J. Haines, and A. L. Goodwin, Journal of the American Chemical Society **134**, 4454 (2012).
[14]     S. J. Hibble, G. B. Wood, E. J. Bilbé, A. H. Pohl, M. G. Tucker, A. C. Hannon, and A. M. Chippindale, Zeitschrift für Kristallographie – Crystalline Materials **225**, 457 (2010).
[15]     S. J. Hibble, S. G. Eversfield, A. R. Cowley, and A. M. Chippindale, Angewandte Chemie International Edition **43**, 628 (2004).
[16]     O. Reckeweg, C. Lind, A. Simon, and J. Salvo, Zeitschrift für Naturforschung B **58**, 155 (2003).
[17]     H. M. Rietveld, Journal of Applied Crystallography **2**, 65 (1969).
[18]     M. J. Cliffe and A. L. Goodwin, Journal of Applied Crystallography **45**, 1321 (2012).
[19]     Y. Liu et al., Physical Chemistry Chemical Physics **26**, 1722 (2024).
[20]     P. Hermet, J. Catafesta, J. L. Bantignies, C. Levelut, D. Maurin, A. B. Cairns, A. L. Goodwin, and J. Haines, The Journal of Physical Chemistry C **117**, 12848 (2013).
[21]     H. Fang, M. T. Dove, and K. Refson, Physical Review B **90**, 054302 (2014).
[22]     C. Romao, M. M. Barsan, I. S. Butler, and D. F. R. Gilson, Journal of Materials Science **45**, 2518 (2010).
[23]     A. L. Goodwin and C. J. Kepert, Physical Review B **71**, 140301 (2005).
[24]     R. Mittal, M. K. Gupta, and S. L. Chaplot, Progress in Materials Science **92**, 360 (2018).
[25]     W. Cochran, Ferroelectrics **35**, 3 (1981).
[26]     G. Lucazeau, Journal of Raman Spectroscopy **34**, 478 (2003).
[27]     G. H. Wolf and R. Jeanloz, Journal of Geophysical Research: Solid Earth **89**, 7821 (1984).
[28]     A. A. Maradudin and A. E. Fein, Physical Review **128**, 2589 (1962).

Supplementary Material for

# Giant Negative Linear Compressibility in Orthorhombic Copper Cyanide

Swayam Kesari[1], Alka B. Garg[2,3], Nilesh P. Salke[1,4*] and Rekha Rao[1,3] *

[1]Solid State Physics Division, [2]High Pressure and Synchrotron Radiation Physics Division, Bhabha Atomic Research Centre, Mumbai 400085, India.

[3] Homi Bhabha National Institute, Anushaktinagar, Mumbai 400094, India

[4] Department of Physics, University of Illinois Chicago, 845 W Taylor St, Chicago 60607, USA

*Corresponding authors
rekhar@barc.gov.in
nilesh@uic.edu

## EXPERIMENTAL DETAILS

A polycrystalline sample of high purity (99.0 %) CuCN (as purchased from Sigma Aldrich) is used in all the experiments. The powder X-ray diffraction (XRD) pattern of the sample matches with the orthorhombic phase ($C222_1$) reported for LT-CuCN [1,2]. Two different spectrographs were used to record Raman spectra in different regions. The excitation wavelength used was 532 nm, ~ 15 mW power. The low-frequency modes in the range 10-100 cm$^{-1}$ are recorded by micro-Raman LabRAM HR-800 Evolution with 50X objective using Ultra Low Frequency (ULF) filter. Modes above 100 cm$^{-1}$ were recorded using a home-built 0.9 m single monochromator, coupled with an edge filter and detected by a cooled CCD. Temperature-dependent Raman spectroscopic measurements from 77 to 293 K were performed using Linkam THMS 600 temperature stage with better temperature stability than ±0.5 K over the sample. Pressure-dependent Raman spectroscopic measurements were performed using a diamond anvil cell (Diacell B-05). To maintain the hydrostatic environment inside the sample chamber, a 16:3:1 methanol−ethanol−water mixture was used as a pressure-transmitting medium (PTM). This PTM remains hydrostatic up to 10.5 GPa then becomes quasi-hydrostatic in the range 10.2-20 GPa.[3] Pressure was measured using ruby fluorescence method.[4,5] At about 10 GPa, the inhomogeneity in the pressure is about 0.2 GPa. The angle dispersive powder x-ray diffraction (ADXRD) measurements at various pressures were recorded at Xpress beam line of Elettra Synchrotron Source, Trieste, Italy. A Mao-Bell type diamond anvil cell was used for pressure generation and the sample with a few grains of copper as *in-situ* pressure marker, along with 16:3:1 methanol-ethanol-water mixture as PTM was loaded inside a stainless-steel gasket. The pressure was calibrated using the equation of state of copper with an accuracy of 0.1 GPa.[6] X-ray beam of wavelength 0.4957 Å collimated to 80 μm was used to collect the powder diffraction data on the MAR345 image plate area detector with a resolution of 100 μm x 100 μm pixel size. Typically, data were collected for an exposure time of 30 seconds at each pressure point. Sample-to-detector distance and various detector orientation parameters were refined with FIT2D software[7] using the diffraction pattern of $CeO_2$. The same software was used to convert two-dimensional diffraction rings from the sample to standard one-dimensional intensity vs. 2θ plots. The Rietveld [8] refinement of the diffraction peaks from the sample and further data analysis were performed with GSAS software.[9] Due to the challenges in refining the C and N positions from the XRD data, we fixed the C and N positions to their reported values [1] while allowing the Cu positions to vary during the refinement.

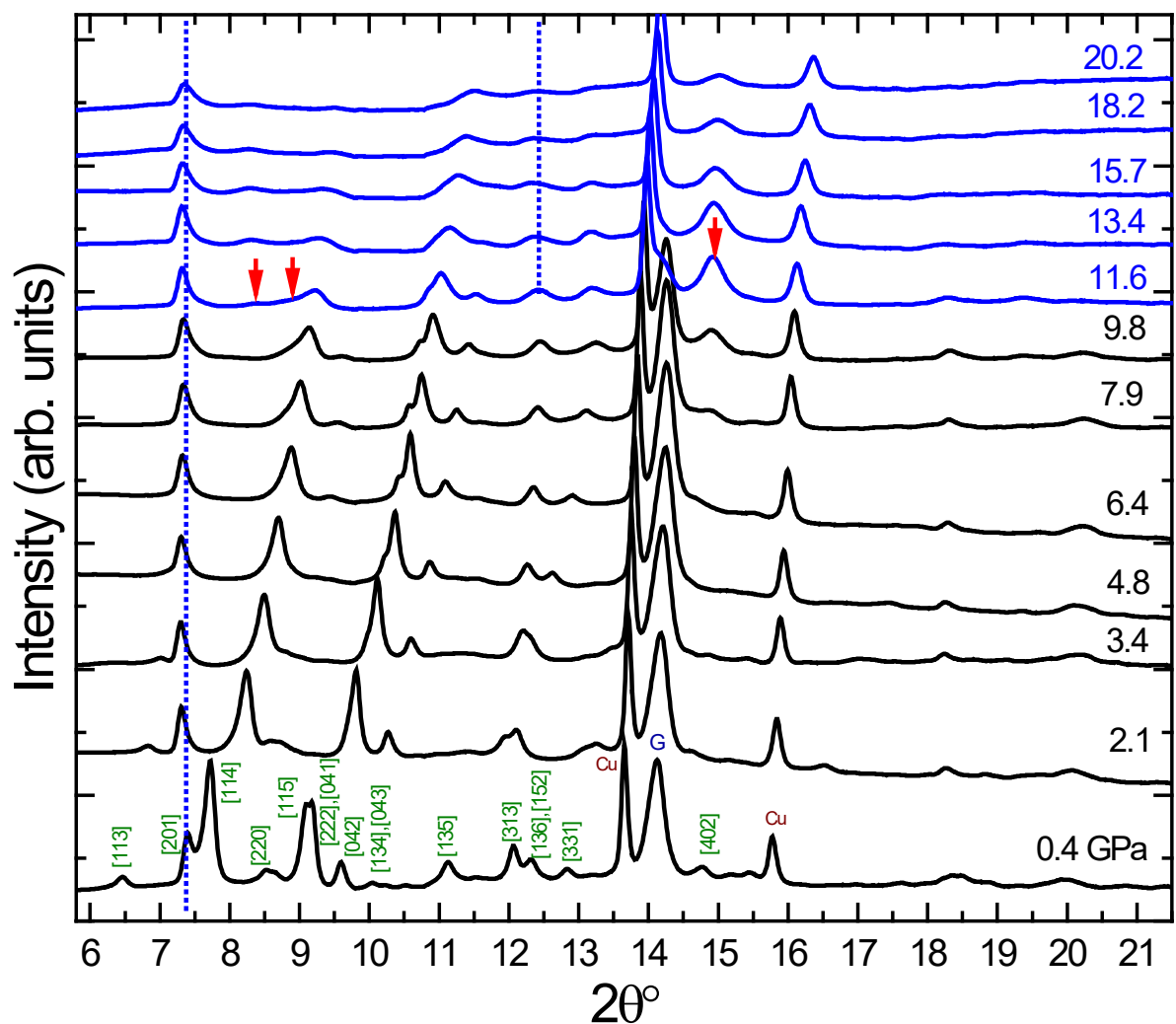

**FIGURE S1:** X-ray diffraction pattern of LT-CuCN at various high pressures. ($\lambda$ = 0.4957 Å). Peaks labeled with Cu indicate reflection peaks for copper, which is used as a pressure marker. Peaks labeled with G indicate the reflection peaks of the gasket material. Vertical dotted blue lines are a guide to see the anomalous movement of reflection peak towards the lower 2θ angles with increasing pressures.

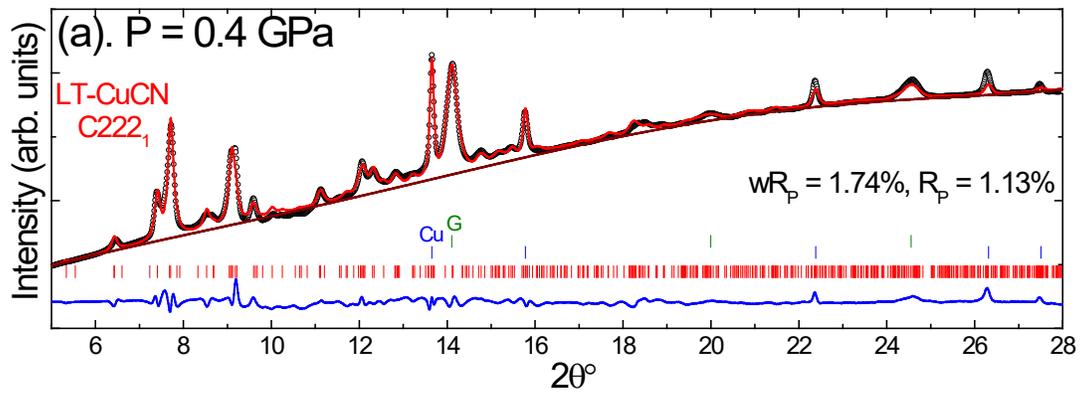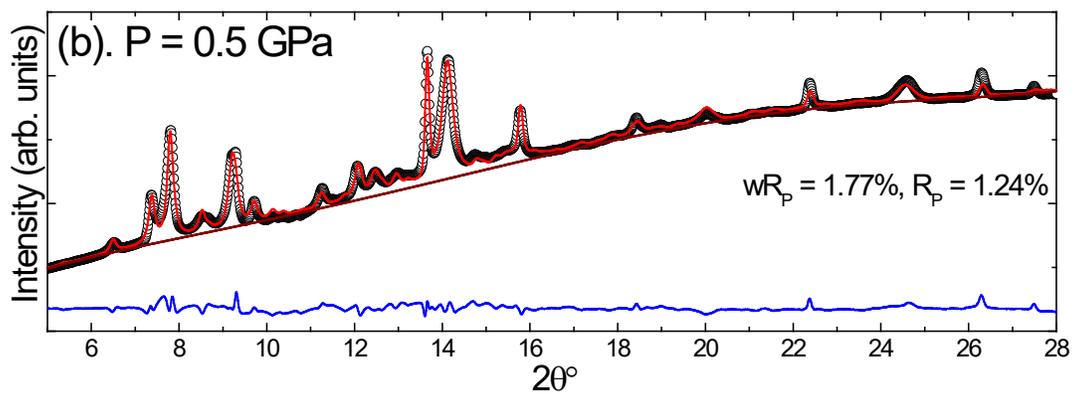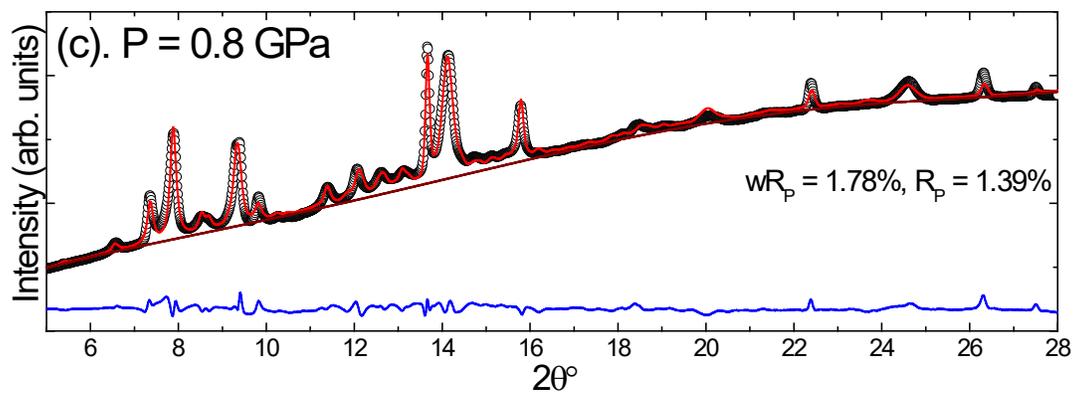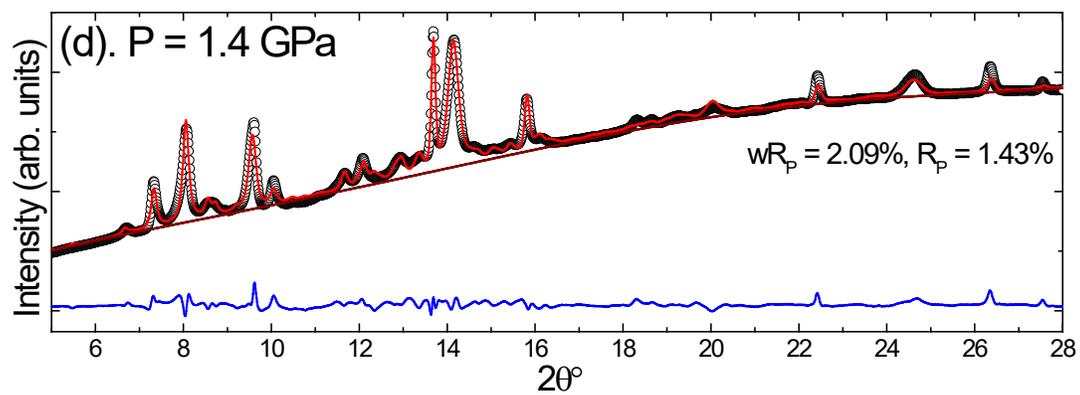

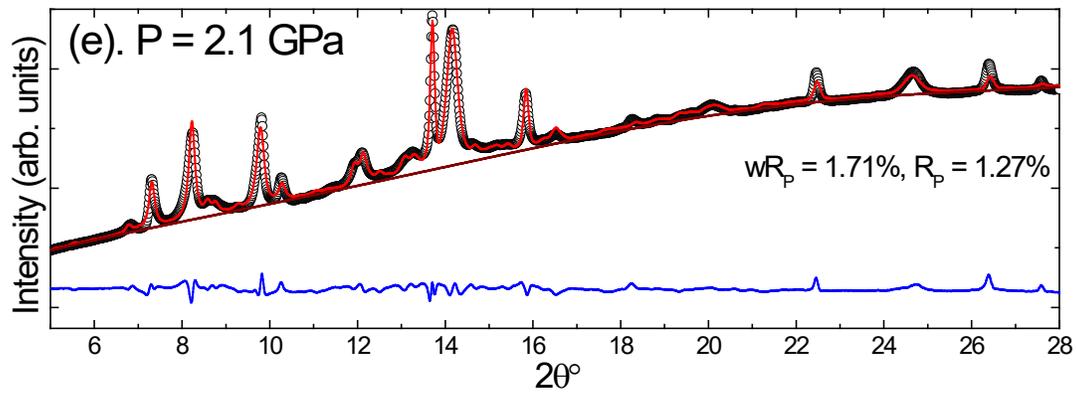

(e). P = 2.1 GPa
wR$_P$ = 1.71%, R$_P$ = 1.27%

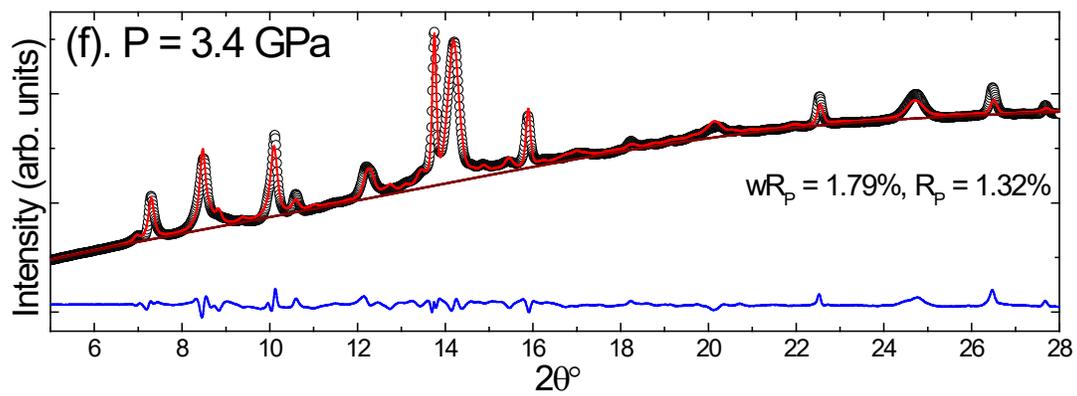

(f). P = 3.4 GPa
wR$_P$ = 1.79%, R$_P$ = 1.32%

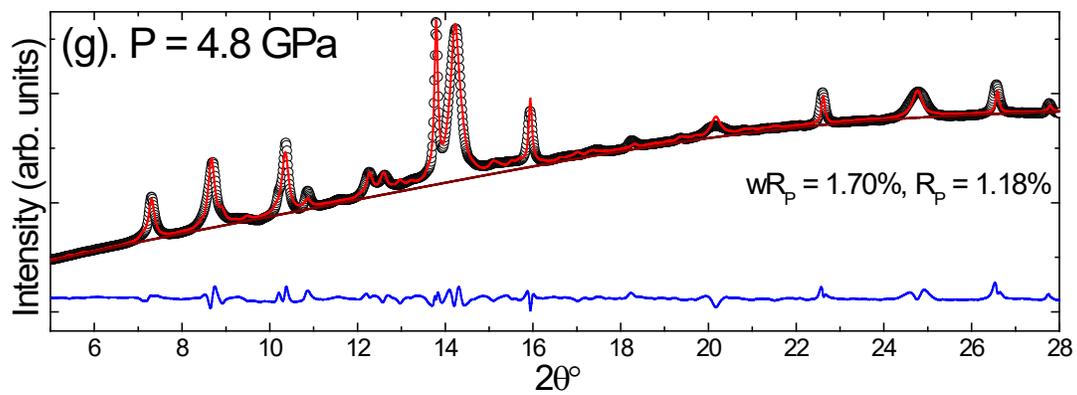

(g). P = 4.8 GPa
wR$_P$ = 1.70%, R$_P$ = 1.18%

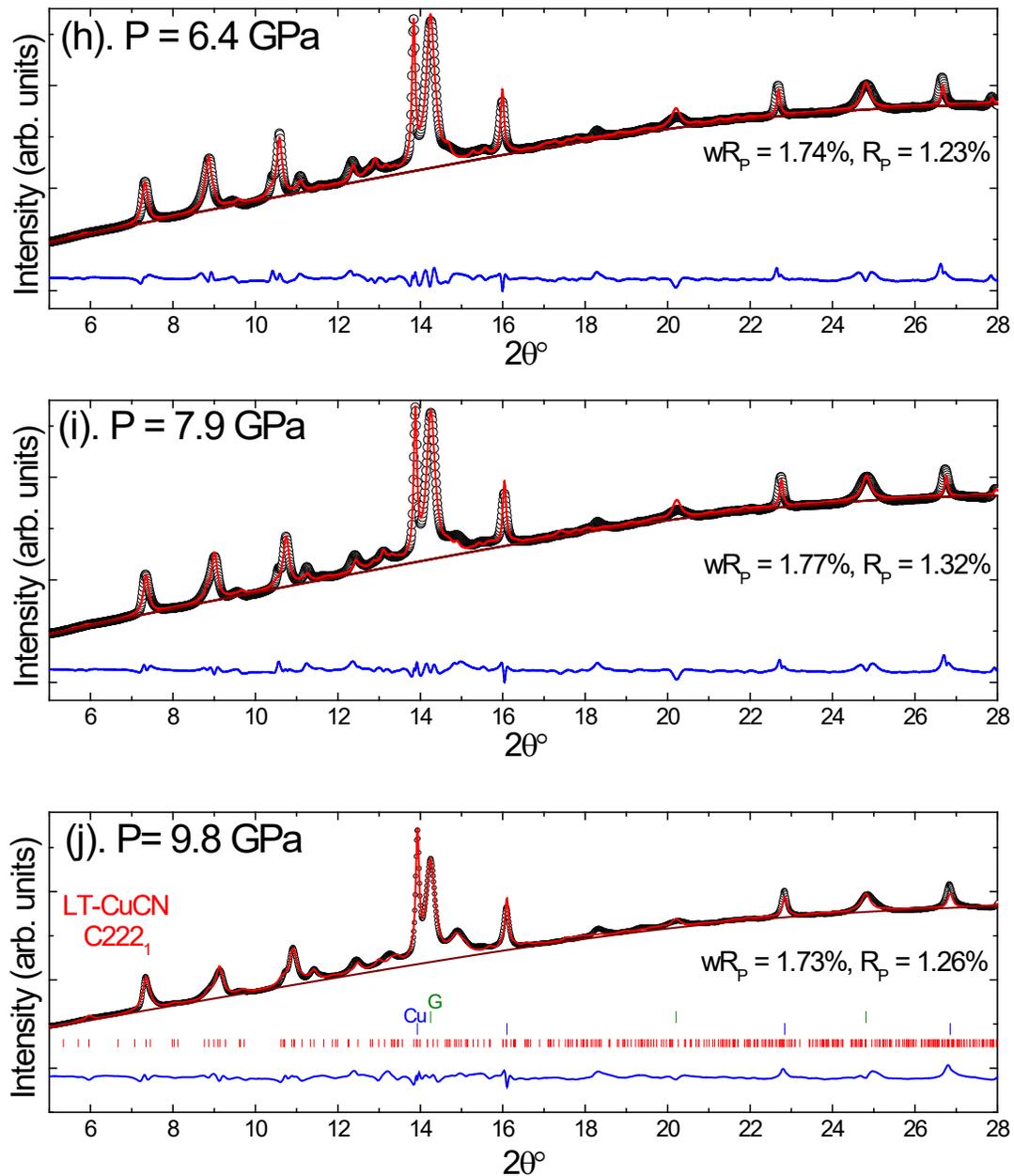

**FIGURE S2:** Rietveld refinement of the XRD pattern of LT-CuCN at a pressure of **(a)** 0.4, **(b)** 0.5, **(c)** 0.8, **(d)** 1.4, **(e)** 2.1, **(f)** 3.4 **(g)** 4.8, **(h)** 6.4, **(i)** 7.9 and **(j)** 9.8 GPa. Peaks labeled with Cu indicate reflection peaks for copper, which is used as a pressure marker. Peaks labeled with G indicate the reflection peaks of the gasket material.

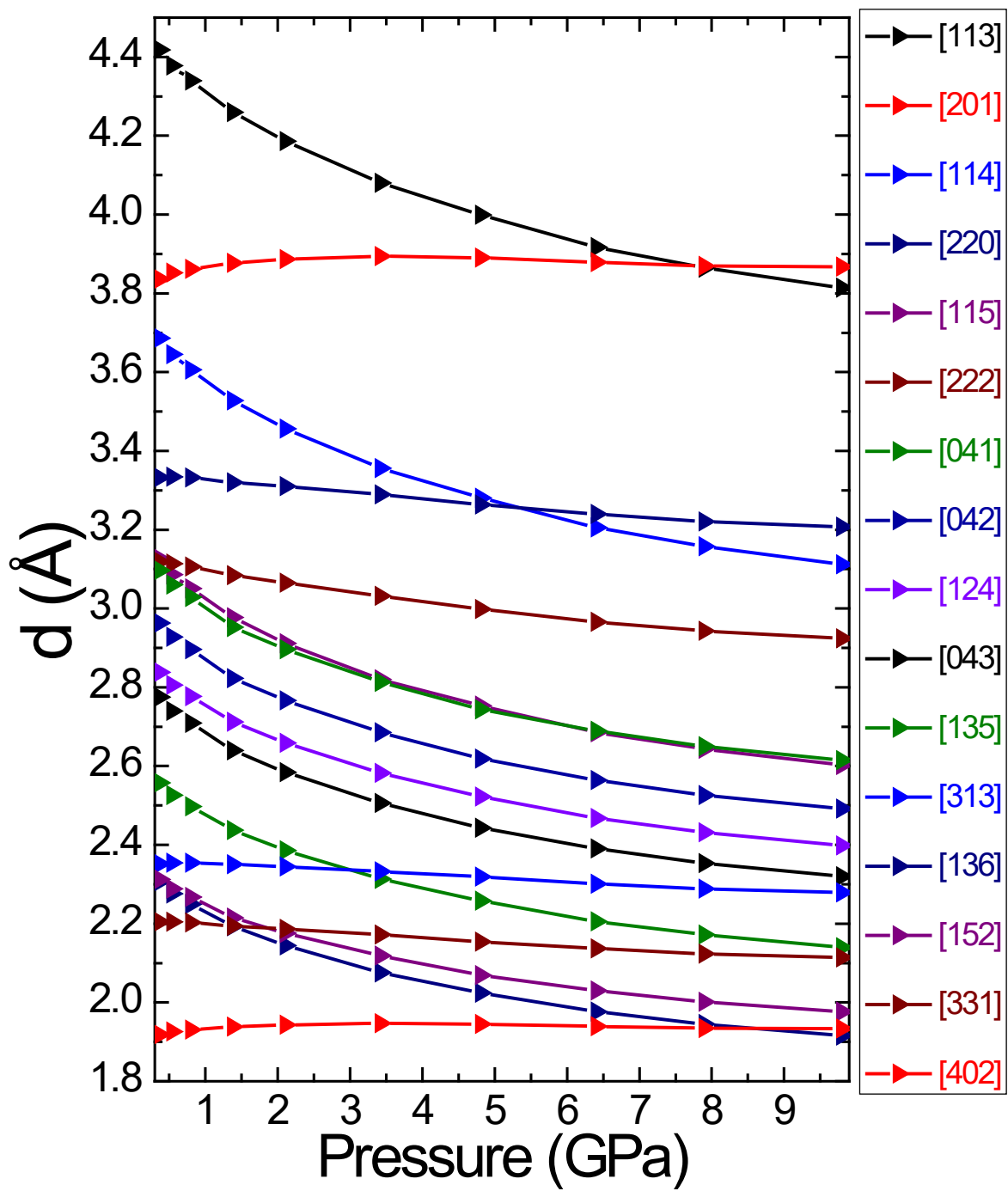

**FIGURE S3:** Variation of inter-planar spacing of prominent lattice planes of LT-CuCN in its orthorhombic structure

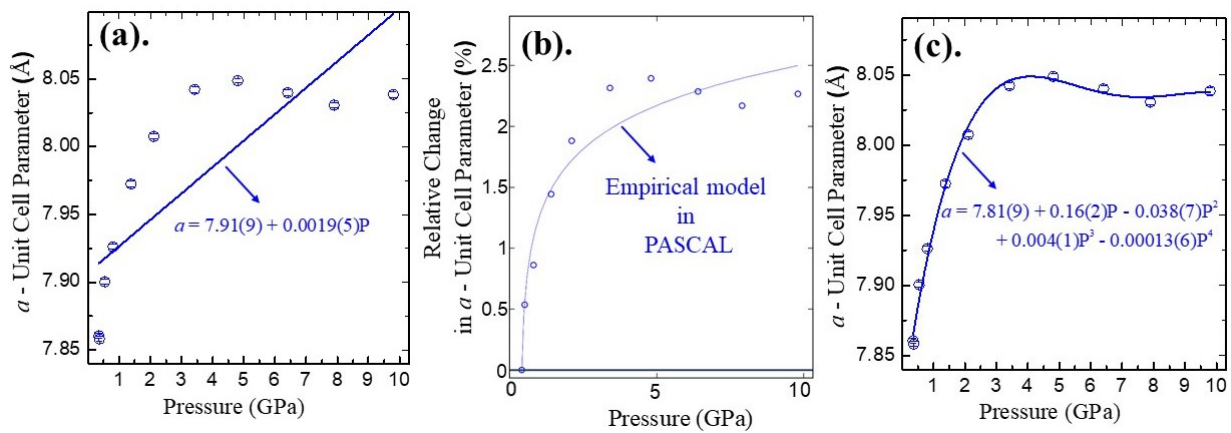

**Figure S4**: Pressure dependence of *a*-parameter. Symbols show experimental *a*-unit cell parameters, Continuous line/curve indicates fitting to (a) Linear, **(b)** Empirical model in PASCAL[10] and **(c)** Polynomial.

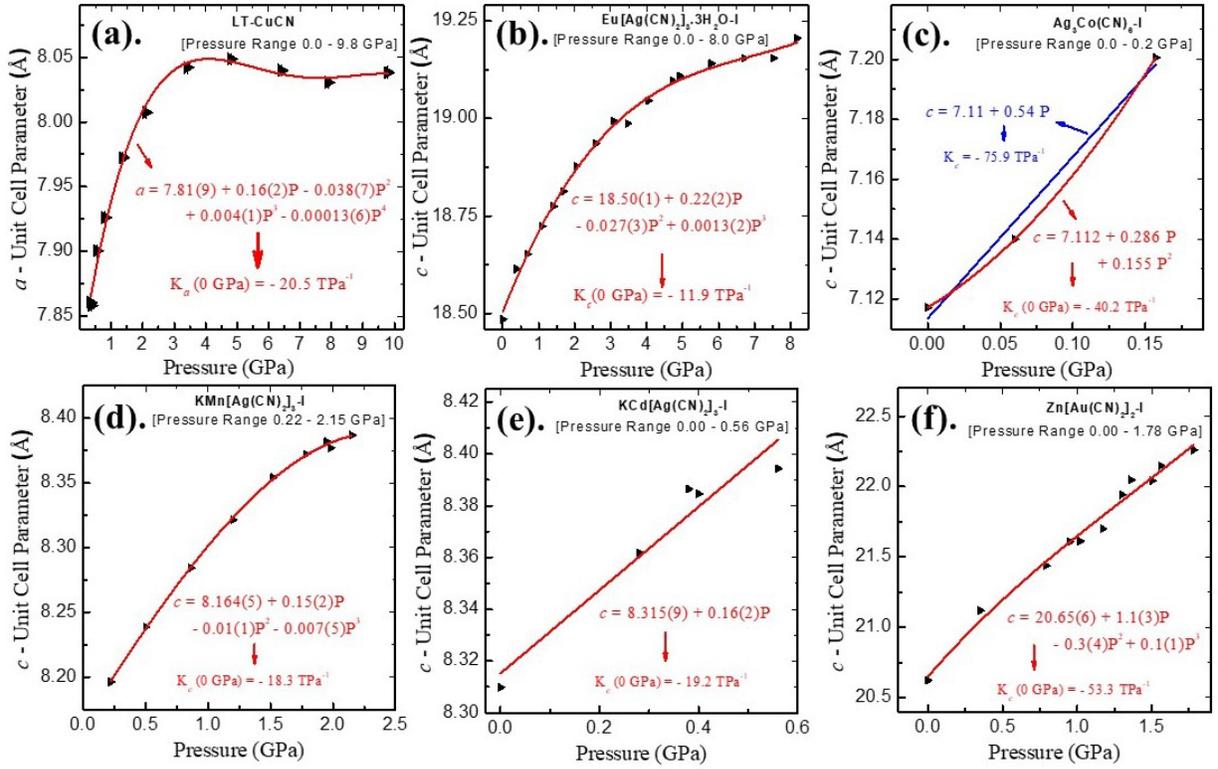

**FIGURE S5:** Pressure variation of NLC lattice parameter of **(a)** LT-CuCN (from present study), **(b)** Eu[Ag(CN)$_2$]3·3H$_2$O (from literature [11]), **(c)** Ag$_3$[Co(CN)$_6$] (from literature [12]), **(d)** KMn[Ag(CN)$_2$]$_3$ (from literature [13]), **(e)** KCd[Ag(CN)$_2$]$_3$ (from literature [14]) and **(f)**. Zn[Au(CN)$_2$]$_2$ (from literature [15]). The continuous curves indicate polynomial fit (to lowest order for giving best fit) and the fitted polynomials are also given. In (c) the linear fit is not a good fit.

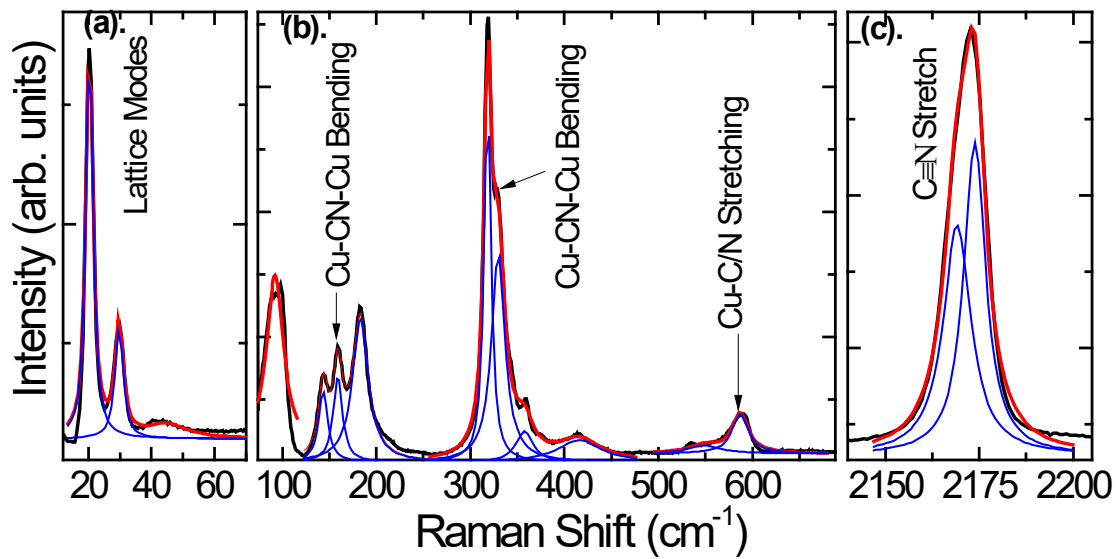

**FIGURE S6:** Raman spectra of LT-CuCN at ambient conditions showing Lorentzian fitting. The assignments are taken from literature.[16]

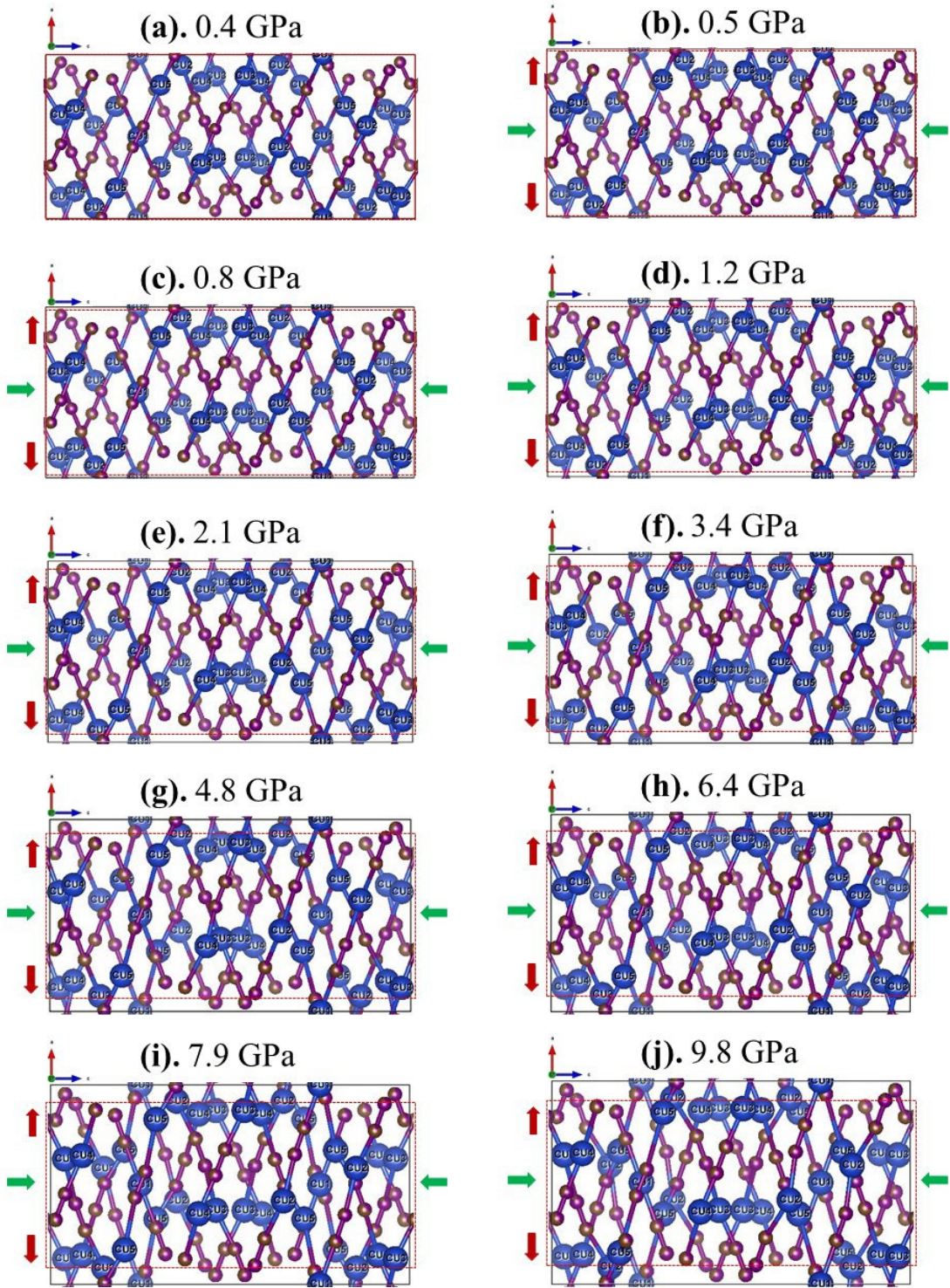

**FIGURE S7 (a)-(j):** Structure of LT-CuCN in the orthorhombic structure at various high pressures viewed in the ac plane.

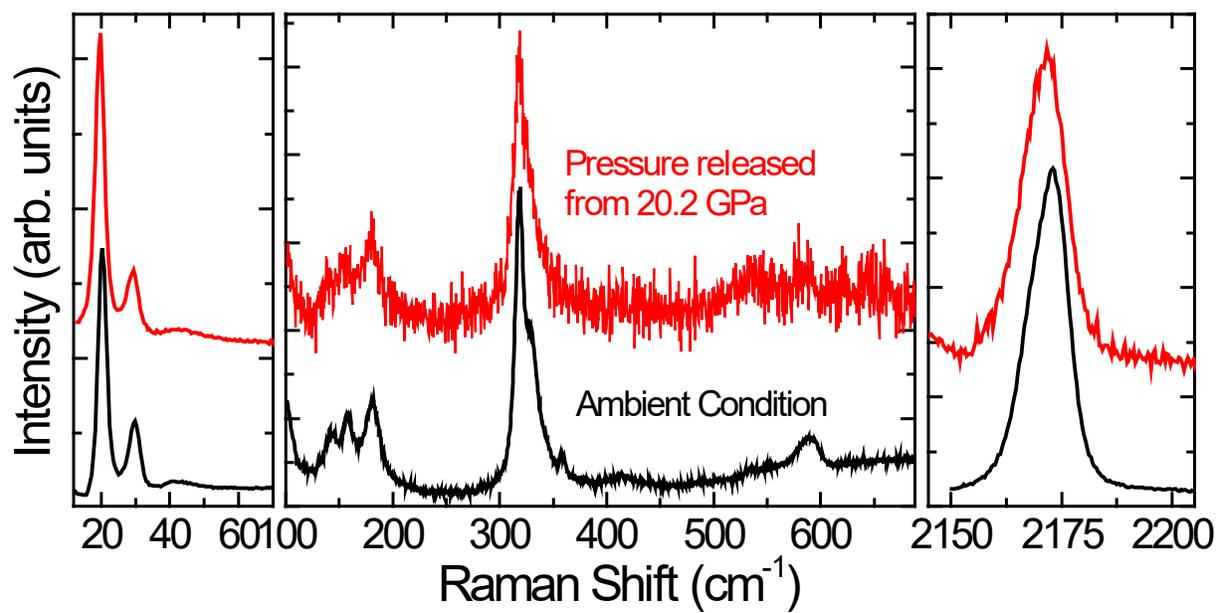

**Figure S8:** Comparison of the Raman spectrum of the fresh and the pressure released LT-CuCN from 20.2 GPa.

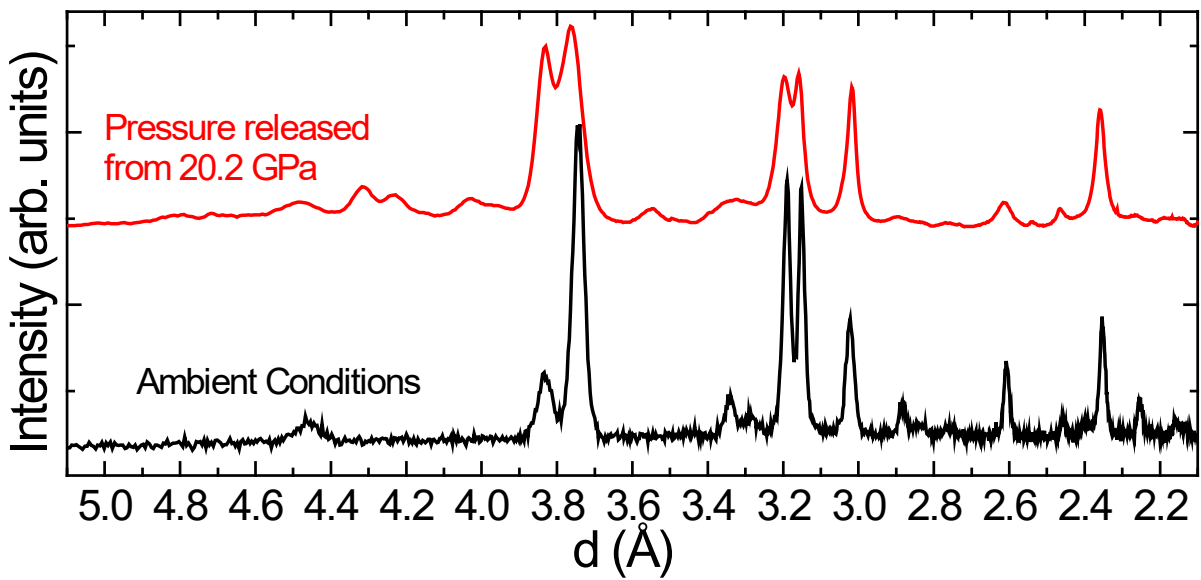

**Figure S9:** Comparison of X-ray diffraction pattern of the fresh and the pressure released LT-CuCN from 20.2 GPa.

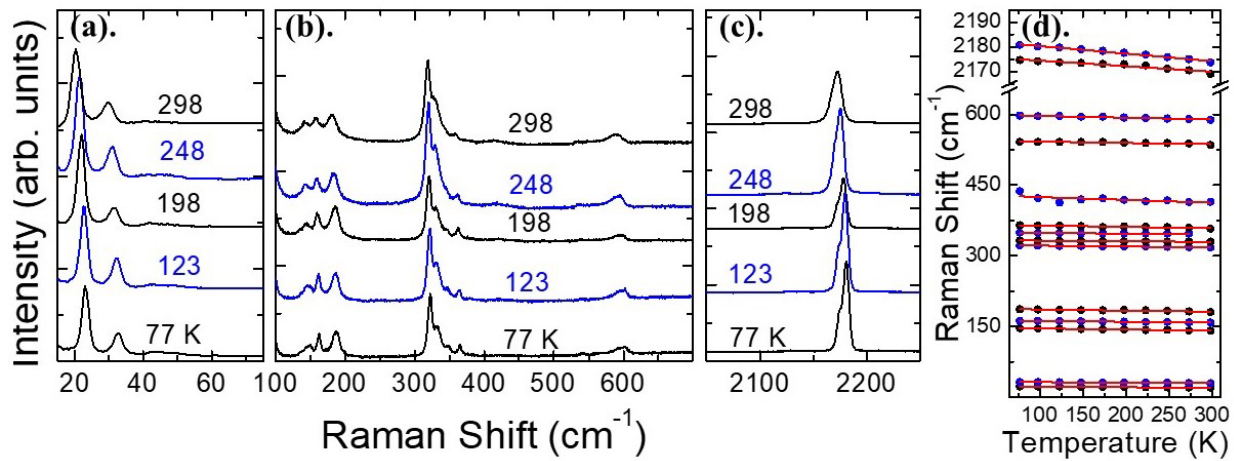

**FIGURE S10:** Raman spectra of LT-CuCN at various temperatures in three regions separated in **(a), (b)** and **(c)**. **(d).** Variation of Raman mode frequencies with temperature. Solid line indicates linear fit to the data.

## TABLE S1: Rietveld refined lattice parameters and fractional coordinates of LT-CuCN at P = 0.4 GPa

a = 7.858(2) Å
b = 12.579(3) Å
c = 17.700(3) Å

|     |   | x        | y       | z       | SOF  | U      |
|-----|---|----------|---------|---------|------|--------|
| Cu1 | 1 | 1.5      | 0.40486 | 0.75    | 1    | 0.0387 |
| Cu2 | 1 | 0.93413  | 0.47564 | 0.63138 | 1    | 0.041  |
| Cu3 | 1 | 0.36866  | 0.35786 | 0.53975 | 1    | 0.0368 |
| Cu4 | 1 | -0.16597 | 0.34657 | 0.42096 | 1    | 0.0387 |
| Cu5 | 1 | -0.68486 | 0.28336 | 0.31037 | 1    | 0.039  |
| N1  | 2 | 1.2851   | 0.4723  | 0.7062  | 0.32 | 0.0325 |
| N2  | 2 | 1.152    | 0.4613  | 0.6803  | 0.68 | 0.0288 |
| N3  | 2 | 0.7297   | 0.4237  | 0.5964  | 0.17 | 0.0354 |
| N4  | 2 | 0.5953   | 0.4103  | 0.5696  | 0.83 | 0.0318 |
| N5  | 2 | 0.1765   | 0.3679  | 0.4861  | 0.37 | 0.0372 |
| N6  | 2 | 0.0466   | 0.3551  | 0.459   | 0.63 | 0.0325 |
| N7  | 2 | -0.3739  | 0.3131  | 0.3747  | 0.3  | 0.0326 |
| N8  | 2 | -0.5068  | 0.3012  | 0.349   | 0.7  | 0.0344 |
| N9  | 2 | -0.9353  | 0.2758  | 0.2645  | 0.5  | 0.0337 |
| C1  | 3 | 1.2851   | 0.4723  | 0.7062  | 0.68 | 0.0325 |
| C2  | 3 | 1.152    | 0.4613  | 0.6803  | 0.32 | 0.0288 |
| C3  | 3 | 0.7297   | 0.4237  | 0.5964  | 0.83 | 0.0354 |
| C4  | 3 | 0.5953   | 0.4103  | 0.5696  | 0.17 | 0.0318 |
| C5  | 3 | 0.1765   | 0.3679  | 0.4861  | 0.63 | 0.0372 |
| C6  | 3 | 0.0466   | 0.3551  | 0.459   | 0.37 | 0.0325 |
| C7  | 3 | -0.3739  | 0.3131  | 0.3747  | 0.7  | 0.0326 |
| C8  | 3 | -0.5068  | 0.3012  | 0.349   | 0.3  | 0.0344 |
| C9  | 3 | -0.9353  | 0.2758  | 0.2645  | 0.5  | 0.0337 |

## TABLE S2: Rietveld Refined lattice parameters and fractional coordinates of LT-CuCN at P = 0.5 GPa

a = 7.900(2) Å
b = 12.433(3) Å
c = 17.410(3) Å

|     |   | x        | y       | z       |   | U      |
|-----|---|----------|---------|---------|---|--------|
| Cu1 | 1 | 1.5      | 0.40252 | 0.75    | 1 | 0.0387 |
| Cu2 | 1 | 0.93269  | 0.48205 | 0.62806 | 1 | 0.041  |
| Cu3 | 1 | 0.36952  | 0.36425 | 0.53643 | 1 | 0.0368 |
| Cu4 | 1 | -0.1742  | 0.34835 | 0.41687 | 1 | 0.0387 |
| Cu5 | 1 | -0.68263 | 0.28213 | 0.31363 | 1 | 0.039  |

Position of N and C atoms are same as in TABLE S1

| TABLE S3 :Rietveld Refined lattice parameters and fractional coordinates of LT-CuCN at P = 0.8 GPa ||||||||
|---|---|---|---|---|---|---|---|
| a = 7.926(2) Å  b = 12.307(4)Å  c = 17.154(3) Å ||||||||
| Cu1 | 1 | 1.5 | 0.40603 | 0.75 | 1 | 0.0387 ||
| Cu2 | 1 | 0.93219 | 0.48146 | 0.63299 | 1 | 0.041 ||
| Cu3 | 1 | 0.38147 | 0.3618 | 0.53479 | 1 | 0.0368 ||
| Cu4 | 1 | -0.17386 | 0.35203 | 0.42001 | 1 | 0.0387 ||
| Cu5 | 1 | -0.67458 | 0.29375 | 0.31135 | 1 | 0.039 ||
| Position of N and C atoms are same as in TABLE S1 ||||||||

| TABLE S 4: Rietveld Refined lattice parameters and fractional coordinates of LT-CuCN at P = 1.4 GPa ||||||||
|---|---|---|---|---|---|---|---|
| a = 7.972(2) Å  b = 12.030(4) Å  c = 16.656(3) Å ||||||||
| Cu1 | 1 | 1.5 | 0.41215 | 0.75 | 1 | 0.0387 ||
| Cu2 | 1 | 0.9367 | 0.48189 | 0.63371 | 1 | 0.041 ||
| Cu3 | 1 | 0.37918 | 0.36593 | 0.53312 | 1 | 0.0368 ||
| Cu4 | 1 | -0.16878 | 0.34918 | 0.42563 | 1 | 0.0387 ||
| Cu5 | 1 | -0.67585 | 0.29363 | 0.30644 | 1 | 0.039 ||
| Position of N and C atoms are same as in TABLE S1 ||||||||

| TABLE S 5: Rietveld Refined lattice parameters and fractional coordinates of LT-CuCN at P = 2.1 GPa ||||||||
|---|---|---|---|---|---|---|---|
| a = 8.007(2) Å  b = 11.773(3) Å  c = 16.209(3) Å ||||||||
| Cu1 | 1 | 1.5 | 0.40329 | 0.75 | 1 | 0.0387 ||
| Cu2 | 1 | 0.93446 | 0.48701 | 0.63849 | 1 | 0.041 ||
| Cu3 | 1 | 0.38054 | 0.36656 | 0.52808 | 1 | 0.0368 ||
| Cu4 | 1 | -0.15943 | 0.34211 | 0.43095 | 1 | 0.0387 ||
| Cu5 | 1 | -0.67683 | 0.29425 | 0.30413 | 1 | 0.039 ||
| Position of N and C atoms are same as in TABLE S1 ||||||||

### TABLE S 6: Rietveld Refined lattice parameters and fractional coordinates of LT-CuCN at P = 3.4 GPa

a = 8.042(2) Å
b = 11.441(4) Å
c = 15.605(4) Å

| | | | | | | |
|---|---|---|---|---|---|---|
| Cu1 | 1 | 1.5 | 0.41229 | 0.75 | 1 | 0.0387 |
| Cu2 | 1 | 0.92639 | 0.49039 | 0.63421 | 1 | 0.041 |
| Cu3 | 1 | 0.38399 | 0.37938 | 0.52104 | 1 | 0.0368 |
| Cu4 | 1 | -0.17196 | 0.33914 | 0.42951 | 1 | 0.0387 |
| Cu5 | 1 | -0.68328 | 0.3031 | 0.2972 | 1 | 0.039 |

Position of N and C atoms are same as in TABLE S1

### TABLE S 7: Rietveld Refined lattice parameters and fractional coordinates of LT-CuCN at P = 4.8 GPa

a = 8.049(2) Å
b = 11.157(2) Å
c = 15.181(3) Å

| | | | | | | |
|---|---|---|---|---|---|---|
| Cu1 | 1 | 1.5 | 0.43334 | 0.75 | 1 | 0.0387 |
| Cu2 | 1 | 0.9198 | 0.48913 | 0.63916 | 1 | 0.041 |
| Cu3 | 1 | 0.37735 | 0.37317 | 0.52588 | 1 | 0.0368 |
| Cu4 | 1 | -0.15949 | 0.33717 | 0.43413 | 1 | 0.0387 |
| Cu5 | 1 | -0.68647 | 0.32105 | 0.30042 | 1 | 0.039 |

Position of N and C atoms are same as in TABLE S1

### TABLE S 8: Rietveld Refined lattice parameters and fractional coordinates of LT-CuCN at P = 6.4 GPa

a = 8.040(2) Å
b = 10.940(3) Å
c = 14.755(3) Å

| | | | | | | |
|---|---|---|---|---|---|---|
| Cu1 | 1 | 1.5 | 0.43379 | 0.75 | 1 | 0.0387 |
| Cu2 | 1 | 0.91257 | 0.47637 | 0.64259 | 1 | 0.041 |
| Cu3 | 1 | 0.37562 | 0.37165 | 0.53574 | 1 | 0.0368 |
| Cu4 | 1 | -0.15501 | 0.34939 | 0.42378 | 1 | 0.0387 |
| Cu5 | 1 | -0.67806 | 0.31592 | 0.29745 | 1 | 0.039 |

Position of N and C atoms are same as in TABLE S1

### TABLE S 9: Rietveld Refined lattice parameters and fractional coordinates of LT-CuCN at P = 7.9 GPa

a = 8.031(2) Å
b = 10.781(3) Å
c = 14.493(4) Å

| | | | | | | |
|---|---|---|---|---|---|---|
| Cu1 | 1 | 1.5 | 0.4603 | 0.75 | 1 | 0.0387 |
| Cu2 | 1 | 0.91888 | 0.44817 | 0.64865 | 1 | 0.041 |
| Cu3 | 1 | 0.37154 | 0.35936 | 0.53931 | 1 | 0.0368 |
| Cu4 | 1 | -0.1458 | 0.36037 | 0.41223 | 1 | 0.0387 |
| Cu5 | 1 | -0.67263 | 0.31037 | 0.29406 | 1 | 0.039 |

Position of N and C atoms are same as in TABLE S1

### TABLE S 10: Rietveld Refined lattice parameters and fractional coordinates of LT-CuCN at P = 9.8 GPa

a = 8.038(2) Å
b = 10.635(4) Å
c = 14.238(4) Å

| | | | | | | |
|---|---|---|---|---|---|---|
| Cu1 | 1 | 1.5 | 0.47618 | 0.75 | 1 | 0.0387 |
| Cu2 | 1 | 0.91453 | 0.45634 | 0.66259 | 1 | 0.041 |
| Cu3 | 1 | 0.36401 | 0.3542 | 0.53386 | 1 | 0.0368 |
| Cu4 | 1 | -0.14375 | 0.36993 | 0.41455 | 1 | 0.0387 |
| Cu5 | 1 | -0.65437 | 0.31676 | 0.31051 | 1 | 0.039 |

Position of N and C atoms are same as in TABLE S1

| TABLE S11: Pressure dependent inter-planar spacing of prominent lattice planes of LT-CuCN in its orthorhombic structure. | | | | | | | | | | |
|---|---|---|---|---|---|---|---|---|---|---|
| Pressure (GPa) | 0.4 | 0.5 | 0.8 | 1.4 | 2.1 | 3.4 | 4.8 | 6.4 | 7.9 | 9.8 |
| $d_{[113]}$ (Å) | 4.418 | 4.378 | 4.34 | 4.259 | 4.186 | 4.08 | 3.999 | 3.917 | 3.865 | 3.814 |
| $d_{[201]}$ (Å) | 3.836 | 3.852 | 3.861 | 3.877 | 3.887 | 3.894 | 3.89 | 3.879 | 3.869 | 3.868 |
| $d_{[114]}$ (Å) | 3.686 | 3.645 | 3.606 | 3.528 | 3.456 | 3.356 | 3.281 | 3.206 | 3.158 | 3.112 |
| $d_{[220]}$ (Å) | 3.332 | 3.334 | 3.332 | 3.32 | 3.311 | 3.29 | 3.264 | 3.239 | 3.22 | 3.207 |
| $d_{[115]}$ (Å) | 3.126 | 3.086 | 3.05 | 2.978 | 2.912 | 2.82 | 2.753 | 2.686 | 2.643 | 2.602 |
| $d_{[222]}$ (Å) | 3.119 | 3.113 | 3.106 | 3.084 | 3.065 | 3.031 | 2.998 | 2.966 | 2.943 | 2.924 |
| $d_{[041]}$ (Å) | 3.096 | 3.06 | 3.028 | 2.952 | 2.896 | 2.813 | 2.743 | 2.689 | 2.65 | 2.615 |
| $d_{[042]}$ (Å) | 2.963 | 2.927 | 2.896 | 2.822 | 2.766 | 2.686 | 2.618 | 2.564 | 2.526 | 2.492 |
| $d_{[124]}$ (Å) | 2.838 | 2.806 | 2.777 | 2.713 | 2.659 | 2.583 | 2.522 | 2.468 | 2.432 | 2.398 |
| $d_{[043]}$ (Å) | 2.775 | 2.74 | 2.709 | 2.639 | 2.585 | 2.506 | 2.443 | 2.39 | 2.354 | 2.32 |
| $d_{[135]}$ (Å) | 2.558 | 2.526 | 2.498 | 2.437 | 2.386 | 2.313 | 2.258 | 2.206 | 2.172 | 2.14 |
| $d_{[313]}$ (Å) | 2.352 | 2.355 | 2.354 | 2.351 | 2.345 | 2.333 | 2.319 | 2.301 | 2.288 | 2.279 |
| $d_{[136]}$ (Å) | 2.306 | 2.276 | 2.249 | 2.193 | 2.144 | 2.076 | 2.025 | 1.976 | 1.945 | 1.915 |
| $d_{[152]}$ (Å) | 2.313 | 2.289 | 2.267 | 2.215 | 2.176 | 2.118 | 2.069 | 2.03 | 2.002 | 1.976 |
| $d_{[331]}$ (Å) | 2.204 | 2.205 | 2.203 | 2.194 | 2.187 | 2.172 | 2.154 | 2.137 | 2.124 | 2.114 |
| $d_{[402]}$ (Å) | 1.918 | 1.926 | 1.931 | 1.938 | 1.943 | 1.947 | 1.945 | 1.939 | 1.935 | 1.934 |

**TABLE S12.** Comparison of axial compressibility of cyanides that shows NLC behaviour in their ambient phase. Linear compressibilities are estimated using polynomial fitting to the reported data [11-15] marked as #, from empirical model in PASCAL [10] marked as & and from linear fitting marked as %.

| Cyanides | Phase | Pressure range of structural stability of the ambient phase (GPa) | Mean Negative Linear Compressibility TPa$^{-1}$ | | Zero Pressure Negative Linear Compressibility TPa$^{-1}$ | Zero Pressure Bulk modulus (GPa) | NTE (Linear Thermal Expansion Coefficient) ($10^{-6}$K$^{-1}$) |
|---|---|---|---|---|---|---|---|
| | | | Reported | Polynomial Fit# | | | |
| **LT-CuCN** | C222$_1$ | 0.0 - 9.8 | No Report | $K_a$ = -10.2 | $K_a$ = -20.5 | 9.2(3) | $α_a$ = -53.8 |
| **Eu[Ag(CN)$_2$]$_3$•3H$_2$O** | P6$_3$/mcm | 0.0 - 8.2 | $K_c$ = -4.2 [11] | $K_c$ = -6.2 | $K_c$ = -11.9# | 16.9(6) | |
| **Ag$_3$Co(CN)$_6$** | P-31m | 0.0 - 0.2 | $K_c$ = -76% [12] | $K_c$ = -43.3 | $K_c$ = -40.2# | 6.5(3) | $α_c$ = −130 |
| **KMn[Ag(CN)$_2$]$_3$** | P312 | 0.0 - 2.8 | $K_c$ = -12& [13] | $K_c$ = -11.2 | $K_c$ = -18.4# | 12.7 | $α_c$ = - 60(3) |
| **KCd[Ag(CN)$_2$]$_3$** | P312 | 0.0 - 0.5 | $K_c$ = -21% [14] | $K_c$ = -19.2 | $K_c$ = -19.2# | <5 | $α_c$ = - 65 |
| **Zn[Au(CN)$_2$]$_2$** | P6$_2$2$_2$ | 0.0 - 1.8 | $K_c$ = -42& [15] | $K_c$ = -40.7 | $K_c$ = -53.3# | 16.7 | $α_c$ = - 57.58(8) |

**TABLE S13.** Pressure dependence and isothermal mode Grüneisen parameters of Raman active modes in LT-CuCN. Isothermal mode Grüneisen parameter $\gamma_{iT} = \frac{B_0}{\omega_i}\left(\frac{d\omega_i}{dP}\right)_T$ is calculated using bulk modulus of LT-Cu(CN)$_2$ which is $B_0 = 9.2$ GPa.

[a] Represents new Raman modes that appeared at high pressure.

| Assignments | Reported [16] | | Present Investigation | | |
|---|---|---|---|---|---|
| | $\omega$ (cm$^{-1}$) | $\left(\frac{\partial \omega}{\partial P}\right)_T$ (cm$^{-1}$/ GPa) | $\omega$ (cm$^{-1}$) | $\left(\frac{\partial \omega}{\partial P}\right)_T$ (cm$^{-1}$/ GPa) | $\gamma_{iT} = \frac{B_0}{\omega_i}\left(\frac{\partial \omega_i}{\partial P}\right)_T$ |
| Lattice Modes | | | 20 | 2.3(2) | 1.05 |
| | | | 30 | 5.4(7) | 1.65 |
| | | | 95 | 1.1(5) | 0.10 |
| Cu-C-N-Cu Bending Motion | | | 141 | -6.6(4) | -0.42 |
| | | | 158 | -7.7(9) | -0.44 |
| | | | 174 [a] | 0.6(2) | 0.03 |
| | | | 181 | 12.1(4) | 0.61 |
| | 315 | 4.9(2) | 318 | 4.4(4) | 0.12 |
| | 326 | 3.9(5) | 330 | 8(1) | 0.23 |
| | | | 346 | | |
| | 358 | -3.3(6) | 357 | 8(3) | 0.21 |
| | | | 414 | -4.1(8) | -0.09 |
| Cu-C/N Stretching Motion | | | 535 | | |
| | 587 | -7.9(8) | 588 | -5.7(8) | -0.09 |
| C-N stretching Motion | 2169 | -4.8(1) | 2169 | -6.3(5) | -0.03 |
| | 2175 | -3.7(1) | 2174 | -5.1(3) | -0.02 |

**TABLE S14.** Raman active modes and their anharmonicity parameters are estimated after combining pressure-dependent and temperature-dependent Raman investigations. Anharmonicity parameters are estimated using $\frac{1}{\omega_i}\left(\frac{\partial \omega_i}{\partial T}\right)_P = -\alpha_V \gamma_{iT} + \frac{1}{\omega_i}\left(\frac{\partial \omega_i}{\partial T}\right)_V$. The thermal expansion coefficient of LT-CuCN $\alpha = 115 \times 10^{-6}$ K$^{-1}$ was estimated using parameters given in the reference [1]. [a] Represents the extrapolated position at 0 GPa for the mode that appears under pressure.

| $\omega_i$ (cm$^{-1}$) | Anharmonicity (x 10$^{-5}$) (K$^{-1}$) | | |
|---|---|---|---|
| | Total $\frac{1}{\omega_i}\left(\frac{\partial \omega_i}{\partial T}\right)_P$ | Implicit $-\alpha \gamma_{iT}$ | Explicit $\frac{1}{\omega_i}\left(\frac{\partial \omega_i}{\partial T}\right)_V$ |
| 20 | -57(3) | -11.7(1) | -45.3 |
| 30 | -40(3) | -18.4(7) | -21.6 |
| 95 | | -1.2(6) | |
| 141 | -16.1(8) | 4.8(3) | -20.9 |
| 158 | -10.9(4) | 5.0(6) | -15.9 |
| 174 [a] | | -0.4(1) | |
| 181 | -13(1) | -6.8(2) | -6.2 |
| 318 | -5.5(3) | -1.4(1) | -4.1 |
| 330 | -4.3(2) | -2.5(3) | -1.8 |
| 346 | -2.6(3) | | |
| 357 | -8(1) | -2.3(9) | -5.7 |
| 414 | -15(6) | 1.0(2) | -16 |
| 535 | -4.8(8) | | |
| 588 | -6.8(5) | 1.0(1) | -7.8 |
| 2169 | -1.0(1) | 0.3(3) | -1.3 |
| 2174 | -1.41(6) | 0.2(1) | -1.61 |

# REFERENCES

[1] S. J. Hibble, S. G. Eversfield, A. R. Cowley, and A. M. Chippindale, Angewandte Chemie International Edition **43**, 628 (2004).
[2] S. J. Hibble, G. B. Wood, E. J. Bilbé, A. H. Pohl, M. G. Tucker, A. C. Hannon, and A. M. Chippindale, Zeitschrift für Kristallographie – Crystalline Materials **225**, 457 (2010).
[3] S. Klotz, J. C. Chervin, P. Munsch, and G. Le Marchand, Journal of Physics D: Applied Physics **42**, 075413 (2009).
[4] D. Errandonea, Y. Meng, M. Somayazulu, and D. Häusermann, Physica B: Condensed Matter **355**, 116 (2005).
[5] G. J. Piermarini, S. Block, J. D. Barnett, and R. A. Forman, Journal of Applied Physics **46**, 2774 (1975).
[6] A. Dewaele, P. Loubeyre, and M. Mezouar, Physical Review B **70**, 094112 (2004).
[7] A. P. Hammersley, S. O. Svensson, M. Hanfland, A. N. Fitch, and D. Hausermann, High Pressure Research **14**, 235 (1996).
[8] H. M. Rietveld, Journal of Applied Crystallography **2**, 65 (1969).
[9] C. Larson A, Report LAUR 86-748, Los Alamos Natonal Laboratory, 121 (1990).
[10] M. J. Cliffe and A. L. Goodwin, Journal of Applied Crystallography **45**, 1321 (2012).
[11] Y. Liu *et al.*, Physical Chemistry Chemical Physics **26**, 1722 (2024).
[12] A. L. Goodwin, D. A. Keen, and M. G. Tucker, Proceedings of the National Academy of Sciences **105**, 18708 (2008).
[13] A. B. Cairns, A. L. Thompson, M. G. Tucker, J. Haines, and A. L. Goodwin, Journal of the American Chemical Society **134**, 4454 (2012).
[14] A. B. Cairns *et al.*, The Journal of Physical Chemistry C **124**, 6896 (2020).
[15] A. B. Cairns *et al.*, Nature Materials **12**, 212 (2013).
[16] C. Romao, M. M. Barsan, I. S. Butler, and D. F. R. Gilson, Journal of Materials Science **45**, 2518 (2010).